\documentclass[lettersize,journal]{IEEEtran}

\usepackage{amsfonts}
\IEEEoverridecommandlockouts
\ifCLASSINFOpdf
\else
\fi
\usepackage{booktabs}
\usepackage{makecell}
\usepackage{graphicx}
\usepackage{subfigure}
\usepackage{cite}
\usepackage{color}
\usepackage[pagebackref=false,breaklinks=false,letterpaper=true,colorlinks,citecolor=blue,linkcolor=blue, anchorcolor=blue, bookmarks=false]{hyperref}
\usepackage{amsmath}
\usepackage{amssymb}
\usepackage{array}
\usepackage{bm}
\usepackage{algorithm}
\usepackage{algorithmic}
\usepackage{pifont}
\newcommand{\et}{\emph{et al.}}

\usepackage{cite}
\usepackage{lettrine} 
\usepackage{orcidlink}

\title{Distributed Multi-Task Learning for Joint Wireless Signal Enhancement and Recognition}
\author{
Hao Zhang, \IEEEmembership{Member, IEEE}, 
Fuhui Zhou, \IEEEmembership{Senior Member, IEEE}, \\
Qihui Wu, \IEEEmembership{Fellow, IEEE}, 
and Chau Yuen, \IEEEmembership{Fellow, IEEE}
\thanks{
This work was supported in part by the National Natural Science Foundation of China under Grant 62222107, in part by the Yangtze River Delta Science and Technology Innovation Community Joint Research (Basic Research) Project under Grant BK20244006, and in part by the Basic Research Projects of Stabilizing Support for Specialty Disciplines under Grant ILF240041A24. (\emph{Corresponding author: Fuhui Zhou}.)

H. Zhang and F. Zhou are with the College of Artificial Intelligence, Q. Wu is with the College of Electronic and Information Engineering, Nanjing University of Aeronautics and Astronautics, Nanjing, 211106, China. They are also with the Key Laboratory of Dynamic Cognitive System of Electromagnetic Spectrum Space of the Ministry of Industry and Information Technology (Nanjing University of Aeronautics and Astronautics), Nanjing, 211106, China (emails: haozhangcn@nuaa.edu.cn, zhoufuhui@ieee.org, wuquhui2014@sina.com)

Chau Yuen is with the School of Electrical and Electronic Engineering, Nanyang Technological University, Singapore 639798, Singapore. (email: chau.yuen@ntu.edu.sg)
}
}

\begin{document}

\maketitle

\begin{abstract}
Wireless signal recognition (WSR) is crucial in modern and future wireless communication networks since it aims to identify the properties of the received signal in a no-collaborative manner. However, it is challenging to accurately classify signals in low signal-to-noise ratio (SNR) conditions and distributed network settings. 
In this paper, we propose a novel distributed multi-task learning framework for joint wireless signal enhancement and recognition (WSER), addressing the crucial need for non-collaborative signal identification in modern wireless networks. Our approach integrates a wireless signal enhancement and recognition network (WSERNet) with FedProx+, an enhanced federated learning algorithm designed for heterogeneous data distributions. 
Specifically, WSERNet leverages an asymmetric convolution block (ACBlock) to capture long-range dependencies in the input signal and improve the performance of the deep learning model. FedProx+ introduces a proximal term to the loss function to encourage the model updates to be closer to the previous model, enhancing the convergence speed and robustness of federated learning. 
Extensive experiments demonstrate the effectiveness of the proposed framework for joint WSER, achieving superior performance compared to state-of-the-art methods under both centralized and distributed settings including independent and identically distributed (IID) and non-IID data distributions.
\end{abstract}


\section{Introduction}

\lettrine[lines=2]{A}{S} wireless communication systems accommodate an ever-increasing number of devices, the demand for efficient spectrum utilization and network management grows exponentially. In this context, wireless signal recognition (WSR) has emerged as a fundamental component in the evolving landscape of modern communication systems. With the electromagnetic spectrum becoming increasingly congested due to the proliferation of wireless devices, the ability to accurately identify, classify, and analyze various signal types has become paramount. Applications of WSR extend far beyond mere signal identification. WSR encompasses critical functions including spectrum management, interference mitigation, security threat detection, and the facilitation of dynamic spectrum access in cognitive radio networks \cite{wang2010advances,zhang2025revolution}. 

Traditional WSR methods can be mainly classified into two categories, namely, likelihood-based (LB) methods and feature-based (FB) methods \cite{dobre2007survey}. LB approaches identify the modulation scheme of received signals by maximizing the likelihood probability under various assumptions. The average likelihood ratio test (ALRT) \cite{huan1995likelihood}, the general likelihood ratio test (GLRT) \cite{panagiotou2000likelihood}, hybrid likelihood ratio test (HLRT) \cite{hameed2009likelihood} methods are three main kinds of LB approaches. Although LB approaches in theory guarantee optimal classification results from the Bayes' sense. However, they have a high computational complexity and require prior knowledge of channel parameters, which is not applicable in practice. FB methods, aim to find better features of the received signals, such as spectral features \cite{nandi1998algorithms}, statistical features  \cite{li2021modulation}, transform features \cite{yuan2004modulation}, and cyclostationarity features \cite{ramkumar2009automatic}. These features are extracted and fed into classifiers to obtain classification results. However, the bias in the estimation of these features has a huge influence on the classification of the modulation scheme. Thus, both LB and FB methods strongly rely on the prior knowledge of the channel information, which is not applicable practically. Moreover, the quality of manually extracted features is limited for WSR. Machine learning (ML) techniques are often utilized with feature-based methods to improve classification accuracy. For example, K-nearest neighbors (KNNs) \cite{zhu2010augmented}, decision trees \cite{luan2022automatic}, support vector machines (SVMs) \cite{wang2009algorithm}, and random forests \cite{triantafyllakis2017phasma} have been used to improve the classification accuracy of WSR. 

Recently, deep learning has been widely applied for wireless communications including WSR. 
Various deep architectures including deep neural networks (DNNs) \cite{niazmand2025joint}, convolutional neural networks (CNNs) \cite{zhang2021novel,zhang2024sswsrnet,yuan2021multiscale}, and recurrent neural networks (RNNs) \cite{ding2022data,huang2020automatic} have achieved remarkable performance in WSR due to their ability to automatically learn features from data. 
DNNs, with their multiple hidden layers, excel at learning complex, non-linear relationships in wireless signals, enabling them to capture intricate signal characteristics that are challenging to extract manually. CNNs have become particularly popular in WSR due to their ability to exploit spatial and temporal correlations in signal data. By applying convolutional filters, CNNs can automatically extract relevant features from raw signal data or time-frequency representations. RNNs, including Long Short-Term Memory (LSTM) \cite{ding2022data} and Gated Recurrent Unit (GRU) \cite{huang2020automatic} variants, are well-suited for processing sequential data in wireless communications. They can capture temporal dependencies in signal patterns, making them valuable for tasks like channel estimation and signal prediction. RNNs have shown particular promise in scenarios involving time-varying channel conditions or when dealing with long sequences of signal data. 
Moreover, techniques including transfer learning \cite{wang2020transfer}, attention mechanisms \cite{zhang2023frequency}, and few-shot learning \cite{zhang2024few} have been introduced to further enhance the performance of deep learning models in WSR tasks. Transfer learning leverages pre-trained models to improve the generalization of deep learning models, enabling them to adapt to new tasks with limited training data. Attention mechanisms focus on relevant signal components, enhancing the model's ability to capture critical features and reduce noise interference. Few-shot learning techniques enable deep learning models to recognize new signal types with minimal training data, offering a more flexible and adaptive approach to WSR tasks. These advanced techniques have the potential to revolutionize WSR by addressing key challenges such as limited data availability, privacy concerns, and communication constraints in wireless networks. 

DL-based methods have revolutionized WSR, demonstrating superior performance compared to traditional approaches. The ability of DL models to automatically extract complex features from raw signal data has led to significant improvements in recognition accuracy and adaptability across various wireless environments. However, these advanced techniques still face considerable challenges, particularly in low signal-to-noise ratio (SNR) conditions and distributed network settings. 
\begin{enumerate}
    \item \emph{Low SNR conditions}: In real-world scenarios, wireless signals are often corrupted by noise, interference, and fading effects, leading to low SNR levels that degrade signal quality. In such challenging environments, DL models may struggle to accurately classify signals due to the presence of background noise and distortion. The performance of DL models is highly dependent on the quality of the training data, making it difficult to achieve robust recognition results in low SNR conditions.
    \item \emph{Distributed network settings}: Due to the mobility of the wireless networks, data samples for WSR are often distributed across multiple edge devices, making it challenging to train DL models in a centralized manner. Moreover, data samples on distributed networks may have differences in distribution, such as independent and identically distributed (IID) and non-IID, which will make model training difficult. 
\end{enumerate}

Traditional WSR approaches treat signal enhancement and recognition as sequential, independent tasks, leading to fundamental limitations that necessitate joint optimization. Conventional enhancement methods optimize for signal reconstruction quality (\emph{e.g.}, minimizing mean squared error (MSE)), which does not align with preserving discriminative features critical for modulation classification. This optimization mismatch can cause enhancement algorithms to inadvertently remove subtle but essential signal characteristics, such as phase transitions in phase-shift keying (PSK) or amplitude variations in quadrature amplitude modulation (QAM), while the sequential pipeline creates an information bottleneck where enhancement errors propagate to recognition without corrective feedback. Moreover, enhancement methods lack knowledge of downstream recognition requirements, making them unable to distinguish between harmful noise and corrupted signal features that still contain valuable modulation information. These fundamental limitations demonstrate that joint optimization is necessary, not merely beneficial, for achieving optimal performance in challenging wireless environments where traditional sequential approaches fail to balance noise reduction with feature preservation.

\begin{figure}
\centering
\includegraphics[width=0.45\textwidth]{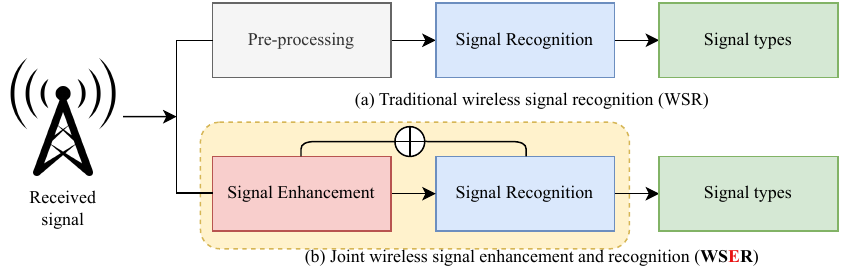}
\caption{Problem definition, (a) traditional wireless signal recognition (WSR), and (b) joint wireless signal enhancement and recognition (\textbf{WS{\color{red}E}R}).}
\label{fig:wser}
\end{figure}

Great efforts have been made to address these challenges, including the development of noise-robust architectures and innovative training strategies. However, there is still a need for more robust and efficient solutions that can adapt to the dynamic and heterogeneous nature of wireless networks. This is mainly because traditional wireless signal recognition (WSR) treats signal enhancement and recognition as two separate tasks, which may lead to suboptimal performance in real-world scenarios. Moreover, signal enhancement still relies on prior knowledge of channel information, which is not applicable in practice. As shown in Fig. \ref{fig:wser} (a), the traditional WSR framework consists of two main components: signal enhancement and signal recognition. The signal enhancement module is responsible for enhancing the received signal by reducing noise and distortion, while the signal recognition module is responsible for recognizing the modulation scheme of the enhanced signal. However, this traditional framework may not be effective in real-world scenarios where the received signal is heavily obscured by background noise, interference, and fading effects. In such challenging environments, DL models may struggle to accurately classify signals due to the presence of noise and distortion. The performance of DL models is highly dependent on the quality of the training data, making it difficult to achieve robust recognition results in low SNR conditions. Thus, we consider the joint wireless signal enhancement and recognition problem, as shown in Fig. \ref{fig:wser} (b). The goal of this problem is to jointly optimize the signal enhancement and recognition tasks to improve the overall performance of WSR in challenging wireless environments. By integrating signal enhancement and recognition into a unified framework, we can leverage the complementary strengths of these tasks to enhance the robustness and accuracy of WSR models. 

Moreover, due to the mobility and heterogeneity of wireless networks, the data samples for WSR are often distributed across multiple edge devices, making it challenging to train a deep learning model in a centralized manner. Federated learning \cite{xu2024distributed,qian2022distributed} is a distributed learning approach that enables multiple edge devices to collaboratively train a deep learning model. Federated learning has been widely used for various applications, such as image classification, speech recognition, and natural language processing. Thus, we try to incorporate federated learning into the joint wireless signal enhancement and recognition problem to address the challenges of data distribution in wireless networks.

In this paper, we propose a novel distributed multi-task learning framework for joint wireless signal enhancement and recognition. The proposed framework leverages the concept of multi-task learning to enable multiple edge devices to collaboratively train a deep learning model for joint wireless signal enhancement and recognition. In particular, each edge device is responsible for collecting and processing local data samples, and then sharing the model updates with a central server for aggregation. The central server aggregates the model updates from all edge devices to generate a global model, which is then shared with all edge devices for further training. By leveraging the distributed nature of multi-task learning, the proposed framework can effectively address the challenges of joint wireless signal enhancement and recognition, such as limited data availability, privacy concerns, and communication constraints. Experimental results on a real-world dataset demonstrate the effectiveness of the proposed framework for joint wireless signal enhancement and recognition. The main contributions of this paper are summarized as follows
\begin{itemize}
    \item We propose a novel distributed multi-task learning framework for joint wireless signal enhancement and recognition based on federated learning. The proposed framework leverages the concept of multi-task learning to enable multiple devices to collaboratively train a deep learning model for joint WSER under both IID and non-IID data distribution. 
    \item we propose a novel asymmetric convolution block (ACBlock) for signal enhancement and recognition to enable multi-scale feature extraction while keeping less computational complexity. The ACBlock is designed to capture long-range dependencies in the input signal and improve the performance of the DL model. Moreover, to facilitate the distributed training process, we propose a novel federated learning algorithm, referred to as \texttt{FedProx+}, which introduces a proximal term to the loss function to encourage the model updates to be closer to the previous model. 
    \item We conduct extensive experiments on both centralized, IID and non-IID distributed settings to evaluate the performance of the proposed framework. The experimental results demonstrate that the proposed framework achieves superior performance compared to state-of-the-art methods for joint WSER.
\end{itemize}

The rest of this paper is organized as follows. Section \ref{sec:related_works} reviews related works. Section \ref{sec:problem_formulation} formulates the problem. Section \ref{sec:method} presents the proposed method. Section \ref{sec:experimental_results} presents experimental results. Section \ref{sec:conclusion} concludes the paper.

\section{Related Works}
\label{sec:related_works}

In this section, we explore approaches related to signal enhancement, signal recognition and federated learning. Firstly, we discuss signal enhancement methods including traditional signal enhancement methods and deep learning-based signal enhancement methods. Then, we present signal recognition methods including likelihood-based (LB) methods, feature-based (FB) methods, and machine learning-based methods. Next, we discuss deep learning-based methods for signal recognition. After that, we introduce federated learning mainly focusing on model aggregation. 

\subsection{Signal Enhancement}
The WSR methods are sensitive to the noise and distortion in the received signals. Therefore, signal enhancement is crucial for improving the performance of WSR methods. Signal enhancement methods aim to improve the quality of the received signals by reducing noise and distortion. Traditional signal enhancement methods can be classified into spatial domain methods and transform domain methods. Spatial domain methods operate directly on the pixel values of the signal, mainly based on filter-based methods \cite{gao2009denoising} and non-local means (NLM) \cite{tracey2012nonlocal}. 
Transform domain methods, on the other hand, operate in a transformed representation of the signal. The most prominent among these is wavelet-based denoising \cite{sendur2002bivariate}, which decomposes the signal into different frequency subbands and applies thresholding to remove noise. Other transform domain techniques include those based on the Fourier transform \cite{chiron2014efficient} and the more recent curvelet and contourlet transform.
Recently, deep learning-based signal enhancement methods have been proposed to improve the performance of WSR methods. These methods leverage the power of deep learning to learn the mapping between noisy and clean signals. For example, DnCNN \cite{zhang2017beyond} is a deep convolutional neural network that has been widely used for image denoising. DnCNN uses a residual learning framework to learn the residual between the noisy and clean signals, enabling it to effectively remove noise from the input signal. 

\subsection{Signal Recognition}
\subsubsection{Likelihood-based (LB) Methods} 
LB approaches are based on the likelihood function calculated for the candidate modulations of the received symbol sequence, where the decision is made by a Bayesian hypothesis testing framework with equally prior probability, \emph{e.g.}, average likelihood ratio test (ALRT) \cite{huan1995likelihood}, generalized likelihood ratio test (GLRT) and hybrid likelihood ratio test (HLRT). 
Huan \et \cite{huan1995likelihood} investigated new likelihood functional-based algorithms for MPSK modulation classification in noisy environments, offering theoretical insights into existing classifiers. 
The work \cite{panagiotou2000likelihood} introduced GLRT and HLRT algorithms for random-phase AWGN channels, demonstrating their superior performance over existing methods, particularly for dense, non-constant envelope constellations. 
The authors in \cite{hameed2009likelihood} explored HLRT-based and quasi-HLRT-based algorithms for AMC, achieving superior performance in terms of classification accuracy. 
Although LB methods are optimal in the Bayesian sense which minimizes the probability of misclassification, they suffer from poor robustness of model mismatches or high computational complexity in practice, especially for the unknown parameters. 

\subsubsection{Feature-based (FB) Methods and Machine Learning-based Methods} 
FB methods aim to find better features of the received signals, such as spectral features, transform features, statistical features, and cyclostationarity features. 
The authors in \cite{nandi1998algorithms} proposed a feature-based method for AMC using three key spectral features derived from the instantaneous amplitude, phase, and frequency of the intercepted signal, achieving an overall success rate of over $96\%$ at an SNR of 15 dB. 
Yuan \et \cite{yuan2004modulation} developed an AMC algorithm using wavelet transform and pattern recognition that achieved efficient performance at 15dB SNR. 
Li \et \cite{li2021modulation} calculated fourth-order cumulants of the received signal including the superposed signal and noise, and then classified the modulation scheme by maximizing the probability density function. 
The work \cite{ramkumar2009automatic} proposed a cyclostationarity-based AMC method for cognitive radio networks to improve spectral sensing reliability and performance. 

Machine learning (ML) techniques are often utilized with FB methods to improve classification accuracy. 
Zhu \et \cite{zhu2010augmented} pioneered the use of genetic programming with a KNN classifier for automatic digital modulation classification, successfully identifying BPSK, QPSK, 16QAM, and 64QAM at 10dB and 20dB SNRs. 
Luan \et \cite{luan2022automatic} presented a decision tree approach for AMC under mixed noise and fading channels, achieving a superior classification accuracy. 
The authors in \cite{wang2009algorithm} proposed a new algorithm based on fourth and sixth-order cumulants and SVM for recognizing six digital modulations, including 2ASK, 4ASK, 8ASK, 4PSK, 8PSK and 16QAM. 
The work \cite{triantafyllakis2017phasma} incorporated a random forest classifier for AMC in their proposed architecture, achieving the classification of various digital and analog modulations under different SNRs. 

\subsection{Deep Learning-based Methods}

Deep learning has been widely used for WSR due to its ability to automatically learn features from data, offering significant advantages over traditional methods that rely on handcrafted features. Various deep learning architectures, including Deep Neural Networks (DNNs), Convolutional Neural Networks (CNNs), and Recurrent Neural Networks (RNNs), have been applied to WSR tasks with promising results. For example, the authors in \cite{liu2017deep} conducted a comprehensive investigation into the use of advanced deep architectures for Automatic Modulation Classification (AMC), a crucial subtask of WSR. Their study encompassed CNNs, residual networks, densely connected networks (DenseNets), and convolutional long short-term deep neural networks (CLDNNs), demonstrating the potential of these architectures in capturing complex signal characteristics.
Building upon this foundation, researchers have explored more sophisticated approaches to enhance WSR performance. In \cite{zhang2020automatic}, a novel dual-stream structure combining CNN and Long Short-Term Memory (LSTM) networks was proposed. This innovative architecture allows for the pairwise interaction of features learned from two distinct deep-learning models, thereby increasing the diversity and richness of the extracted features. 
Moreover, techniques including transfer learning \cite{wang2020transfer}, attention mechanisms \cite{zhang2023frequency}, and few-shot learning \cite{zhang2024few} have been introduced to further enhance the performance of deep learning models in WSR tasks. 
Wang \et \cite{wang2020transfer} proposed a transfer learning-based semi-supervised AMC method that combines convolutional auto-encoders and neural networks to effectively classify modulation signals with limited labeled data in MIMO systems. The work \cite{zhang2023frequency} introduced a multi-spectral attention mechanism with the discrete cosine transform (DCT) for learning-based frequency component selection in automatic modulation classification, combining hand-crafted features and deep learning to outperform existing methods. 
The authors in \cite{zhang2024few} developed a novel few-shot AMC method combining neural architecture search and knowledge transfer to achieve high classification accuracy with limited labeled data, outperforming competitors in complex electromagnetic environments.

\subsection{Federated Learning}
Federated learning is a distributed learning approach that enables multiple edge devices to collaboratively train a deep learning model. Federated learning has been widely used for various applications, such as image classification, speech recognition, and natural language processing. 
FedAvg \cite{zhang2025federated} is a popular federated learning algorithm that aggregates the model updates from all edge devices to generate a global model. 
FedProx \cite{li2020federated} is an extension of FedAvg that introduces a proximal term to the loss function to encourage the model updates to be closer to the previous model. FedProx has been shown to improve the convergence speed and robustness of federated learning.

\section{Problem Formulation}
\label{sec:problem_formulation}

\subsection{Signal Model}
Consider a single-carrier, single-input single-output (SISO) signal model with constant signal bandwidth and signal length. To simulate the dataset for the deep learning approach, we present a simplified yet accurate model. The received baseband signal $r(t)$ for a wireless transmission affected by channel and hardware impairments is given by
\begin{equation}
r(t) = \exp{(j2\pi f_{\text{err}}t+\theta_{\text{err}})} \sum_{l}h(l) s(t-l-\zeta_{\text{err}}) + w(t),
\end{equation}
where $h$ is the channel impulse response, $s$ denotes the modulated symbol up-sampled and pulse shaped by a root-raised-cosine (RRC) filter, $T_s$ is the sampling rate, $f_{\text{err}}$, $\theta_{\text{err}}$, and $\zeta_{\text{err}}$ denote frequency offset, phase offset, and timing error, respectively. $w$ is the additive white Gaussian noise (AWGN) with zero mean and variance $\sigma^2$. 


The continuous-time received signal is then sampled and quantized to obtain discrete-time complex-valued raw signal symbols. These raw signal symbols can be expressed in terms of their in-phase (I) and quadrature (Q) components as
\begin{equation}
    x[n] = \text{Re}(r[n]) + j\text{Im}(r[n]),
\end{equation}
where $x[n]$ represents the complex-valued raw signal symbol at discrete time index $n$, and $r[n]$ is the sampled version of $r(t)$.

\begin{figure}[!t]
\centering
\includegraphics[width=0.5\textwidth]{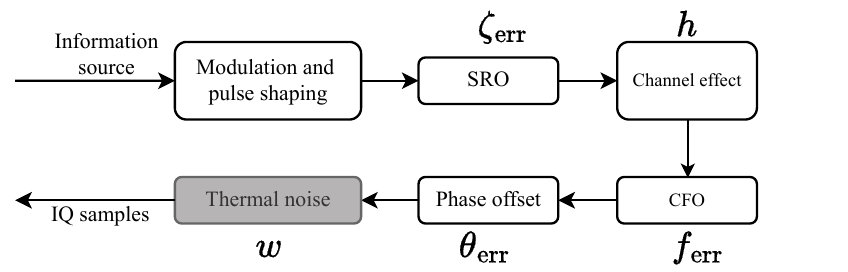}
\caption{Block diagram of the signal model \cite{sathyanarayanan2023rml22}.}
\label{fig:signal}
\end{figure}

\subsection{Problem Statement}

Let $x[n]$ be the observed signal, $s^*[n]$ the signal with channel and hardware effects, and $w[n]$ the noise component
\begin{equation}
    x[n] = s^*[n] + w[n].
\end{equation}
The objective is to estimate $\hat{s}(t)$, such that
\begin{equation}
    \hat{s}[n] \approx s^*[n],
\end{equation}
and the goal is to minimize the error between the enhanced signal $\hat{s}[n]$ and the signal $s^*[n]$, which can be expressed as
\begin{equation}
    \mathbb{E}[|s^*[n] - \hat{s}[n]|^2],
\end{equation}
where $\mathbb{E}$ denotes the expectation operator. 

After the signal enhancement operation, the enhanced signal $\hat{s}[n]$ is then fed into a deep learning model for signal recognition. The goal of signal recognition is to identify the modulation scheme of the received signal based on the enhanced signal $\hat{s}[n]$. 
Given a set of $M$ known signal classes $\{C_1, C_2, ..., C_m\}$, and an enhanced signal $\hat{s}[n]$, the goal is to find a classification function $f(\hat{s}[n])$ such that
\begin{equation}
    f(\hat{s}[n]) = \arg\max_{i\in{1,...,M}} P(C_i|\hat{s}[n]),
\end{equation}
where $P(C_i|\hat{s}[n])$ is the probability that $\hat{s}[n]$ belongs to class $C_i$.

\section{Proposed Method}
\label{sec:method}
This section presents the algorithm and system design of the proposed distributed multi-task learning for joint signal enhancement and recognition based on federated learning. 
We first introduce the system model and the wireless signal enhancement and recognition network (WSERNet). Then, we present the proposed \texttt{FedProx+} algorithm for distributed learning based on federated learning. Finally, we describe the training process and the model deployment process.

\subsection{System Model}

\begin{figure}
\centering
\includegraphics[width=0.45\textwidth]{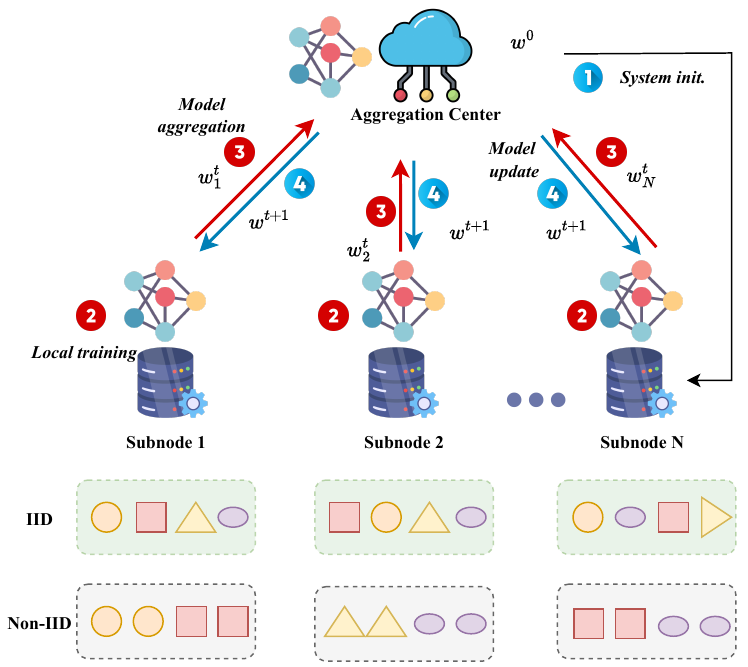}
\caption{System model of wireless signal enhancement and recognition (WSER) in a distributed network.}
\label{fig:system_model}
\end{figure}

In this paper, we propose a distributed system model for wireless signal enhancement and recognition (WSER) implemented across a multi-node network based on federated learning. As illustrated in Fig. \ref{fig:system_model}, the architecture comprises multiple subnodes, each operating an autonomous sub-network model and maintaining a distinct signal sample database. During the training phase, we employ a federated learning approach wherein one node is dynamically designated as the aggregation center from the pool of edge nodes. This aggregation node assumes the critical role of parameter aggregation and output harmonization, facilitating collaborative learning across the network. To ensure uniformity and compatibility, each sub-node is equipped with an identical sub-network model based on our proposed WSENet architecture. The training process is conducted in a distributed manner, which can be given as follows. 
\begin{enumerate}
    \item \emph{System initialization}: The system is initialized with a set of hyperparameters, including the learning rate, batch size, and model architecture. The system parameters are initialized based on the requirements of the training process and the characteristics of the dataset. The system is also initialized with a set of quality criteria to evaluate the performance of the trained model, such as accuracy, convergence rate, and stability. 
    \item \emph{Local training}: The local models are initialized with random parameters $w_0$, and each sub-node independently trains its sub-network model using local signal samples. The training process involves optimizing the model parameters to minimize the joint loss function $\mathcal{L}$ based on the local dataset. The local model parameters $w^t_1$ are updated iteratively using stochastic gradient descent (SGD) or other optimization algorithms, where $t$ denotes the training round, and $n$ is the subnode number. Then, the trained local model parameters $w^t_n$ are shared with the aggregation center for model aggregation.
    \item \emph{Model aggregation}: The aggregation center receives the local model updates from all subnodes and aggregates them to generate a global model. The aggregation process involves combining the local model updates using a weighted average or other aggregation methods. The global model parameters $w^{t}$ are then shared with all subnodes for further training.
    \item \emph{Model update}: The global model parameters $w^{t}$ are shared with all subnodes for further training. The subnodes update their local models using the global model parameters $w^{t}$ and continue the training process. This iterative process continues until the convergence criteria are met, and the global model achieves satisfactory performance.
\end{enumerate}

\begin{figure*}
\centering
\includegraphics[width=0.95\textwidth]{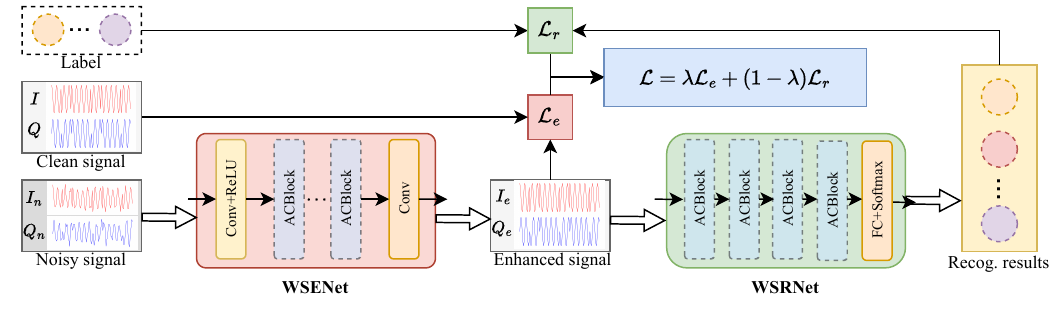}
\caption{Architecture of wireless signal enhancement and recognition network (\textbf{WSERNet}).}
\label{fig:e2e}
\end{figure*}

\subsection{Wireless Signal Enhancement and Recognition Network (WSERNet)}
As shown in Fig. \ref{fig:e2e}, we propose a novel wireless signal enhancement and recognition network (WSERNet) for joint wireless signal {\color{blue}enhancement} and recognition. The WSERNet consists of two main components, namely the wireless signal enhancement network (WSENet), and the wireless signal recognition network (WSRNet). The WSENet is responsible for enhancing the received signal by reducing noise and distortion. The WSRNet is responsible for recognizing the modulation scheme of the enhanced signal. Both these two networks are built based on the proposed asymmetric convolution block (ACBlock). The ACBlock is designed to capture long-range dependencies in the input signal and improve the performance of the DL model. The WSENet and WSRNet are trained end-to-end using a multi-task learning strategy to jointly optimize the signal enhancement and recognition tasks. The WSERNet is designed to be lightweight and efficient, making it suitable for deployment on resource-constrained edge devices. The details of ACBlock, WSENet and WSRNet are described in the following subsections.


\begin{figure}
\centering
\subfigure[]{\includegraphics[width=0.95\linewidth]{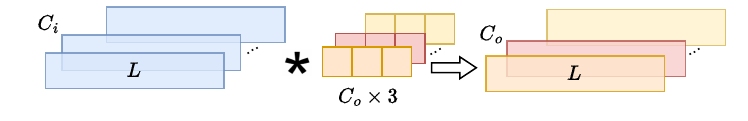}}
\subfigure[]{\includegraphics[width=0.95\linewidth]{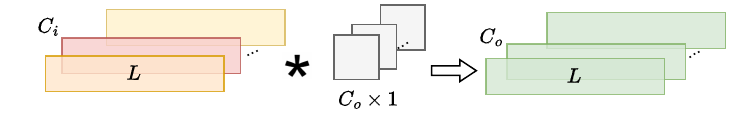}}
\subfigure[]{\includegraphics[width=0.95\linewidth]{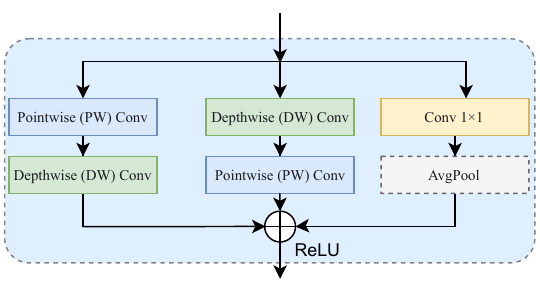}}
\DeclareGraphicsExtensions.
\caption{The architecture of the proposed asymmetric convolution block(ACBlock), (a) depthwise convolution, (b) pointwise convolution, (c) ACBlock.}
\label{fig:acblock}
\end{figure}

\subsubsection{Asymmetric Convolution Block (ACBlock)}
The ACBlock is an innovative 1D convolutional module designed to enhance model expressiveness and computational efficiency. This module comprises three parallel convolutional branches, namely,  pointwise-depthwise convolution branch, depthwise-pointwise convolution branch, and residual branch. 
The pointwise-depthwise convolution branch first applies a $1\times 1$ pointwise convolution for channel reduction and feature aggregation, followed by a depthwise convolution for feature extraction. 
The depthwise-pointwise convolution branch reverses this order, using a depthwise convolution first, then a $1\times 1$ convolution for channel fusion. 
These two branches draw inspiration from the inverted residual structure in MobileNet \cite{sandler2018mobilenetv2} and Efficient \cite{tan2021efficientnetv2}, but with improvements and extensions. The residual employs a $1\times 1$ convolution to preserve original feature information and enable cross-channel information interaction. 
Each convolutional operation is followed by a Batch Normalization layer to improve model stability and convergence speed. When the stride is greater than 1, the residual includes an average pooling layer for downsampling to maintain spatial consistency with other branches. 
The innovation of ACBlock lies in its fusion of multiple convolutional operations, where depthwise convolutions improve computational efficiency, $1\times1$ convolutions enable cross-channel information interaction, and the parallel structure enhances model expressiveness. This design allows ACBlock to capture features at different scales and of different types while maintaining relatively low computational complexity. 
In the forward pass, the outputs of the three branches are combined through element-wise addition to generate the final feature map. This fusion method allows the network to learn the optimal combination of different convolutional operations, further increasing the model's flexibility and adaptability. 
Overall, ACBlock provides an efficient and expressive basic building block for 1D convolutional neural networks, with the potential to improve model performance in various sequence processing tasks.

Given an input tensor $X \in \mathbb{R}^{C_{i} \times L}$, where $C_{i}$ is the number of input channels and $L$ is the feature length, the output $Y \in \mathbb{R}^{C_{o} \times L'}$ of the ACBlock can be expressed as
\begin{equation}
    Y = \mathrm{ReLU}(F_{PD}(X) + F_{DP}(X) + F_{R}(X)),
\end{equation}
Where $F_{PD}(X)$, $F_{DP}(X)$, and $F_{R}(X)$ are the outputs of the pointwise-depthwise, depthwise-pointwise, and residual branches, respectively, which are defined as follows
\begin{subequations}
\begin{align}
F_{PD}(X) &= \mathrm{BN}(\mathrm{DWConv}(\mathrm{BN}(\mathrm{PWConv}(X)))), \label{eq:DG} \\
F_{GD}(X) &= \mathrm{BN}(\mathrm{PWConv}(\mathrm{BN}(\mathrm{DWConv}(X)))), \label{eq:GD} \\
F_{R}(X) &= \mathrm{DS}(\mathrm{BN}(Conv_{1\times1}(X))),
\end{align}
\end{subequations}
where $Conv_{1\times1}$ denotes a $1\times1$ convolution, $\mathrm{PWConv}$ is the pointwise convolution, $\mathrm{DWConv}$ denotes a depthwise separable convolution, $\mathrm{BN}$ denotes a Batch Normalization layer, and $\mathrm{DS}$ denotes an average pooling layer. The output $Y$ is the element-wise sum of the three branches, which is then passed through a ReLU activation function to introduce non-linearity. The output $Y$ is then passed to the next layer in the network for further processing.

\subsubsection{Wireless Signal Enhancement Network (WSENet)} 
DnCNN (Denoising Convolutional Neural Network) is a pioneering deep learning model for image denoising, introduced by \cite{zhang2017beyond}. It employs a convolutional neural network architecture typically consisting of 17 to 20 layers, utilizing residual learning to predict noise rather than clean images directly. This approach, combined with batch normalization and ReLU activation, enables DnCNN to handle various noise types and intensities without prior knowledge of noise levels. The model's effectiveness stems from its ability to learn end-to-end mappings between noisy and clean images, outperforming traditional methods, especially at higher noise levels. 
As shown in Fig. \ref{fig:e2e}, we follow the structure of DnCNN and change the basic convolutional blocks with the proposed ACBlock. The WSENet is designed to enhance the received signal by reducing noise and distortion. The WSENet is trained end-to-end using a multi-task learning strategy to jointly optimize the signal enhancement and recognition tasks. The WSENet is designed to be lightweight and efficient, making it suitable for deployment on resource-constrained edge devices. The structure of WSENet is shown in Table \ref{tab:wsenet_structure}. 

\begin{table}[!ht]
\centering
\caption{The Structure of WSENet}
\begin{tabular}{clcc}
\toprule
\textbf{Layer} & \textbf{Type} & \textbf{Output} & \textbf{Parameters} \\
\midrule
- & input & (2, 128) & 0 \\
\midrule
1 & Conv1D (32, 3, 1) + ReLU & (32, 128) & 256 \\
\midrule
2-16 & ACBlock & (32, 128) & 83,520 \\
\midrule
17 & Conv1D (32, 3, 1) & (2, 128) & 192 \\
\midrule
total parameters & - & - & 83,968 \\
\bottomrule
\end{tabular}
\label{tab:wsenet_structure}
\end{table}

\subsubsection{Wireless Signal Recognition Network (WSRNet)} 
The WSRNet is a deep-learning model designed for wireless signal recognition. The WSRNet consists of a series of convolutional layers followed by a fully connected layer and a softmax layer. The convolutional layers are designed to extract features from the input signal, while the fully connected layer and softmax layer are designed to classify the input signal into one of the $K$ modulation schemes. The WSRNet is trained end-to-end using a multi-task learning strategy to jointly optimize the signal enhancement and recognition tasks. The WSRNet is designed to be lightweight and efficient, making it suitable for deployment on resource-constrained edge devices. The structure of WSRNet is given in Table \ref{tab:wsrnet_structure}.

\begin{table}[h]
\centering
\caption{The Structure of WSRNet}
\begin{tabular}{clcc}
\toprule
\textbf{Layer} & \textbf{Type} & \textbf{Output} & \textbf{Parameters} \\
\midrule
- & input & (2, 128) & 0 \\
\midrule
Conv1 (ACBlock1D) & 2 & (64, 128) & 3,776 \\
\midrule
Conv2 (ACBlock1D) & 64 & (128, 64) & 26,048 \\
\midrule
Conv3 (ACBlock1D) & 128 & (256, 32) & 100,224 \\
\midrule
Conv4 (ACBlock1D) & 256 & (512, 16) & 395,392 \\
\midrule
FC & 512 & (512, $K$) & 5,120 \\
\midrule
Softmax & - & (512, $K$) & 0 \\
\midrule
total parameters & - & - & 530,560 \\
\bottomrule
\end{tabular}
\label{tab:wsrnet_structure}
\end{table}

\subsubsection{Loss Function}
The loss function for the joint wireless signal enhancement and recognition task is defined as the sum of the signal enhancement loss and the signal recognition loss. The signal enhancement loss is defined as the mean squared error (MSE) between the enhanced signal and the clean signal. The signal recognition loss is defined as the cross-entropy loss between the predicted modulation scheme and the ground truth modulation scheme. The total loss is defined as the sum of the signal enhancement loss and the signal recognition loss. The total loss is minimized during training to optimize the signal enhancement and recognition tasks jointly. The loss function is defined as follows
\begin{equation}
    \mathcal{L} = \lambda \mathcal{L}_{\text{e}} + (1-\lambda)\mathcal{L}_{\text{r}},
\end{equation}
where $\mathcal{L}_{\text{e}}$ is the signal enhancement loss, $\mathcal{L}_{\text{r}}$ is the signal recognition loss, $\lambda$ is the weight parameter, 
and $\mathcal{L}$ is the total loss. The signal enhancement loss is defined as
\begin{equation}
    \mathcal{L}_{\text{e}} = \frac{1}{N}\sum_{i=1}^{N}\|s_i - \hat{s}_i\|_2^2,
\end{equation}
where $s_i$ is the clean signal, $\hat{s}_i$ is the enhanced signal, and $N$ is the number of training samples. The signal recognition loss is defined as
\begin{equation}
\mathcal{L}_{\text{r}} = -\frac{1}{N}\sum_{i=1}^{N}\sum_{m=1}^{M}y_{i,m}\log(\hat{y}_{i,m}),
\end{equation}
where $y_{i,m}$ is the ground truth label for the $i$-th sample and the $m$-th modulation scheme, $\hat{m}_{i,m}$ is the predicted probability of the $i$-th sample belonging to the $m$-th modulation scheme, and $M$ is the number of modulation schemes. The total loss is minimized during training to optimize the signal enhancement and recognition task jointly.

\subsection{Distributed Learning based on Federated Learning}

To address the challenges of distributed learning in WSER, we propose a federated learning-based approach that enables multiple edge devices to collaboratively train a deep learning model, referred to \texttt{FedProx+}.  
FedProx+ enhances the original FedProx algorithm for federated learning by introducing client-specific and dynamically adjusted proximal term weights. This approach addresses the limitations of treating all clients equally in heterogeneous environments. FedProx+ initializes each client with an individualized proximal term coefficient (($\mu_k$)) based on its data characteristics. Throughout the training process, these coefficients are dynamically adjusted using an adaptive function that considers both local and global model performance. This mechanism allows the algorithm to better balance local optimization with global model consistency, potentially improving convergence and performance in diverse federated learning scenarios. By adapting to each client's unique characteristics and evolving performance, FedProx+ aims to enhance the robustness and efficiency of federated learning across heterogeneous client devices and data distributions.

As shown in Algorithm \ref{alg:enhanced_fedprox}, FedProx+ operates over $T$ rounds, involving a server and $N$ client devices. The algorithm begins with the server initializing device-specific proximal term coefficients ($\mu_k$) based on each device's data characteristics, as well as setting the initial global model parameters ($w_0$). 
In each round, the server randomly selects a subset of $K$ devices, with each device $k$ having a selection probability of $p_k$. The current global model parameters are then distributed to the selected devices. Each chosen device performs local optimization to find an updated model ($w^{(t+1)}_k$) that minimizes a proximal objective function. This function combines the device's local objective ($F_k$) with a proximal term that encourages similarity to the global model. The optimization is performed to a $k$-inexact minimizer, allowing for computational flexibility.
After local computation, each device calculates a performance metric ($P^{(t+1)}_k$), such as accuracy, and sends this along with its updated model parameters back to the server. The server then computes a global performance metric ($P^{(t+1)}_{global}$) using a test dataset. A key feature of FedProx+ is its adaptive mechanism: the server updates each device's proximal term coefficient ($\mu_k$) using an adaptive function that considers both the device's local performance and the global model's performance. AdaptiveMu is a function that adjusts the proximal term coefficient based on the ratio of the device's local performance to the global performance.


\begin{algorithm}
    \caption{The proposed \texttt{FedProx+} algorithm.}
    \label{alg:enhanced_fedprox}
    \begin{algorithmic}[1]
    \REQUIRE $K$, $T$, $\mu_0$, $\gamma$, $w_0$, $N$, $p_k$, $k = 1, \dots, N$
    \ENSURE The global model $\bm{w_T}$
    \STATE \textbf{Server executes:}
    \STATE \quad Initialize $\mu_k$ for each device $k$ based on its data characteristics
    \STATE \quad Initialize the global model $\bm{w_0}$
    \STATE \quad \textbf{for} round $t = 0, \dots, T - 1$ \textbf{do}
    \STATE \quad \quad Select a subset $S_t$ of $K$ devices at random (each device $k$ is chosen with probability $p_k$)
    \STATE \quad \quad Send $\bm{w_t}$ to all chosen devices
    \STATE \quad \quad \textbf{for} each chosen device $k \in S_t$ \textbf{do}
    \STATE \quad \quad \quad Find $\bm{w_k^{(t+1)}}$ which is a $\gamma_k$-inexact minimizer of:
    \STATE \quad \quad \quad \quad $\arg\min_{\bm{w}} h_k(\bm{w}; \bm{w_t}) = F_k(\bm{w}) + \frac{\mu_k}{2}|\bm{w} - \bm{w_t}|^2$
    \STATE \quad \quad \quad Compute local performance metric $P_k^{(t+1)}$ (e.g., accuracy)
    \STATE \quad \quad \quad Send $\bm{w_k^{(t+1)}}$ and $P_k^{(t+1)}$ back to the server
    \STATE \quad \quad \textbf{end for}
    \STATE \quad \quad Server computes global performance metric $P_{\text{global}}^{(t+1)}$ on test data
    \STATE \quad \quad \textbf{for} each device $k$ \textbf{do}
    \STATE \quad \quad \quad $\mu_k = \text{AdaptiveMu}(P_k^{(t+1)}, P_{\text{global}}^{(t+1)}, \mu_0)$
    \STATE \quad \quad \textbf{end for}
    \STATE \quad \textbf{end for}
    \STATE \textbf{return} The global model $\bm{w_T}$
\end{algorithmic}
\end{algorithm}


\section{Experimental Results}
\label{sec:experimental_results}
\subsection{Dataset}

We utilize a modulation classification dataset generated by the RML22 dataset generation code \cite{sathyanarayanan2023rml22}. This is an updated version of the popular RML16 dataset \cite{o2016convolutional}, providing a more realistic and carefully designed signal model parameterization. The modulated signals of the generated dataset are simulated in a Rayleigh fading channel environment with additive white Gaussian noise and variable delay spreads, Doppler shift, sample rate offset, center frequency offset and phase offset. The generated RML22 dataset consists of a total of 10 different modulation forms, 8 digital (BPSK, QPSK, 8PSK, PAM4, QAM16, QAM64, GFSK and CPFSK) and 2 analogs (WBFM and AM-DSB) modulation forms, where each modulation form has 21 levels of uniformly distributed SNRs from -20 dB to 20 dB in 2 dB steps. The dataset consists of $1,260,000$ sample examples. Each sample example is composed of 128 samples in length for two channels I/Q that can be represented as $2\times 128$ samples. The parameters used in the RML22 dataset generation are shown in Table \ref{tab:rml22_parameters}. The dataset is generated using a set of parameters that are carefully selected to ensure the realism and diversity of the generated signals. The dataset is designed to be used for training and testing deep learning models for wireless signal recognition tasks.

\begin{table}[t]
\centering
\caption{Parameters Used in RML22 Dataset Generation}
\begin{tabular}{p{2cm}p{5cm}}
\toprule
\textbf{Description} & \textbf{RML22 Values} \\
\midrule
Signal & Samples per symbol = 2, Sample rate = 30 kHz \\
 & Center frequency = 1 GHz, Clock rate = 100MHz \\
\midrule
Modulation & Roll-off factor = 0.35, CPFSK modulation index = 0.5, \\
 & GFSK BW time product = 0.3, GFSK sensitivity = 1.57, \\
 & WBFM max. freq. deviation = 75 kHz \\
\midrule
Dataset & Num. frames per mod. per SNR = 2000, Frame length = 128, \\
 & Frame duration = 4.5 ms \\
\midrule
Fading & 3GPP fading model ETU70, \\
 & Filter tap magnitudes = [$-1, -1, -1, 0, 0, -3, -5, -7$] dB, \\
 & Filter tap delays = [0, .05, .12, .2, .23, .5, 1.6, 2.3, 5] ns, \\
 & Num. of taps = 8, Max. freq. dev (Doppler) = 70 Hz, \\
 & Num. of sinusoids = 8 \\
\midrule
Clock effect & XO, LO and SRO max. freq. deviation: 5 Hz, 500 Hz, 50 Hz \\
 & XO, LO and SRO std dev per sample = $10^{-4}$, $10^{-2}$, $10^{-3}$ \\
 & XO to LO scaling = 100, XO to clock rate scaling = 10 \\
\midrule
AWGN & $-20$ to 20 dB in steps of 2 dB \\
\bottomrule
\end{tabular}
\label{tab:rml22_parameters}
\end{table}

In this work, we select a subset of the RML22 dataset for training and testing, with an SNR range of $-6$ dB to $14$ dB in $2$ dB steps, as in our previous work \cite{zhang2024sswsrnet}. 
A total number of $220,000$ samples are selected, where $176,000$ samples are used for training and $44,000$ samples are used for testing. The training set is distributed across multiple edge devices, and the testing set is used to evaluate the performance of the proposed method.

\subsection{Experimental Setup}
Experiments are conducted on a workstation with an AMD Ryzen 9 7900X3D CPU, 64GB RAM, and an NVIDIA GeForce RTX 4070 Ti GPU, and the system is running Ubuntu 22.04. Code is implemented in Python and the PyTorch deep learning framework \cite{paszke2019pytorch}. 

For centralized training, a batch size of $256$ is used, and the model is trained for $100$ epochs with a learning rate of $0.01$. The SGD optimizer is used with a momentum of $0.9$ and a weight decay of $0.0005$. The model is trained end-to-end using the proposed WSERNet architecture. The model is evaluated on the testing set to measure the classification accuracy. 
For distributed training, the proposed \texttt{FedProx+} algorithm is used with a batch size of $64$ and a learning rate of $0.002$. 
The model is trained for $100$ communication rounds with $5$ devices for the small-scale IID distributed setting and $200$ communication rounds with $25$ devices (randomly selected from $50$ distributed devices) for the large-scale IID distributed setting. 
As for the Non-IID distributed setting, the model is trained for $300$ communication rounds with $50$ devices (randomly selected from $100$ distributed devices). 
A comprehensive summary of the experimental parameters and settings is provided in Table \ref{tab:experimental_parameters}.

\begin{table}[h]
\centering
\caption{Key Experimental Parameters and Settings}
\begin{tabular}{ll}
\toprule
\textbf{Parameter} & \textbf{Value} \\
\midrule
Dataset & RML22 subset (-6 to 14 dB, 220K samples) \\
Input format & (2, 128) I/Q samples \\
Train/Test split & 176K / 44K (80\% / 20\%) \\
\midrule
Centralized training & Batch=256, LR=0.01, Epochs=100 \\
Federated training & Batch=64, LR=0.002, $\lambda$=0.3 \\
Communication rounds & 100 (IID-5), 200 (IID-25), 300 (non-IID-50) \\
\midrule
WSENet & 15 ACBlocks, 32 channels \\
WSRNet & 4 ACBlocks [64,128,256,512] + FC \\
Optimizer & SGD (momentum=0.9, decay=0.0005) \\
\midrule
Hardware & RTX 4070 Ti, PyTorch, Ubuntu 22.04 \\
\bottomrule
\end{tabular}
\label{tab:experimental_parameters}
\end{table}

\subsection{Performance Evaluation under Centralized Conditions}
\begin{figure}[!ht]
\centering
\includegraphics[width=0.45\textwidth]{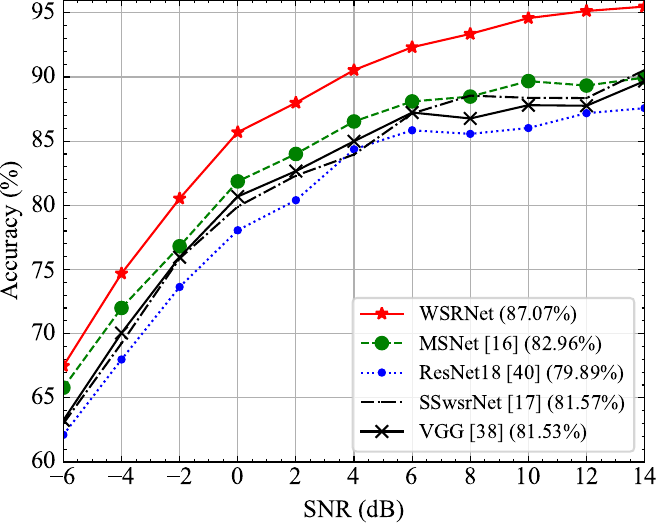}
\caption{Performance comparison with the state-of-the-art models including VGG \cite{o2016convolutional}, ResNet \cite{o2018over}, MSNet \cite{zhang2021novel}, and SSwsrNet \cite{zhang2024sswsrnet} under centralized conditions.}
\label{fig:center_sr}
\end{figure}

To present the effectiveness of the proposed WSRNet for wireless signal recognition tasks, we first evaluate the performance of the proposed method under centralized conditions, as shown in Fig. \ref{fig:center_sr}. The performance of the proposed WSRNet is compared with the state-of-the-art models including VGG \cite{o2016convolutional}, ResNet \cite{o2018over}, MSNet \cite{zhang2021novel}, and SSwsrNet \cite{zhang2024sswsrnet}. 
The results show that the proposed WSRNet achieves the best classification accuracy of $87.07\%$, outperforming the state-of-the-art models. Moreover, it consistently outperforms the other models across all SNR levels, showing a steep increase in accuracy as SNR improves, particularly in the $-4$ to $4$ dB range. MSNet \cite{zhang2021novel} (green dashed line) ranks second with $82.96\%$ accuracy, following a similar trend to WSRNet but with slightly lower performance throughout the SNR range. 
The remaining three models, ResNet \cite{he2016deep} (blue dotted line), SSwsrNet \cite{zhang2024sswsrnet} (black dash-dot line), and VGG [4] (black line with x markers), show comparable performance to each other. Their accuracies range from $79.89\%$ to $81.57\%$, with slight variations in their relative positions across different SNR levels.
All models exhibit a general trend of improved accuracy as SNR increases, with the most significant gains occurring between $-4$ dB and $6$ dB. 
The proposed WSRNet consistently outperforms the other models, demonstrating its superior performance in WSR tasks under centralized conditions.

\begin{table}[h]
\centering
\caption{Computational Complexity and Performance Comparison}
\begin{tabular}{lccc}
\toprule
\textbf{Model} & \textbf{Parameters} & \textbf{FLOPs} & \textbf{Accuracy} \\
 & & \textbf{(M)} & \\
\midrule
VGG \cite{o2016convolutional}  & 36,422,410 & 104.28 & 79.89\% \\
ResNet \cite{o2018over} & 3,849,482 & 43.906 & 81.57\% \\
MSNet \cite{zhang2021novel} & 74,955 & 3.5 & 82.96\% \\
SSwsrNet \cite{zhang2024sswsrnet} & 90,915 & 4.9 & 80.78\% \\
\midrule
\textbf{WSRNet (ours)} & \textbf{530,560} & \textbf{16.6} & \textbf{87.07\%} \\
\textbf{WSERNet (ours)} & \textbf{614,528} & \textbf{18.2} & \textbf{89.60\%} \\
\bottomrule
\end{tabular}
\label{tab:complexity_comparison}
\end{table}

As shown in Table \ref{tab:complexity_comparison}, we compare the computational complexity and performance of the proposed WSRNet and WSERNet with the state-of-the-art models. Our proposed WSERNet demonstrates exceptional computational efficiency while achieving superior performance. With only 614,528 parameters and 18.2M FLOPs, WSERNet achieves the highest accuracy (89.60\%) among all compared methods, representing dramatic improvements over conventional approaches: 9.71\% accuracy gain with 59× fewer parameters than VGG, and 8.03\% improvement with 6.3× fewer parameters than ResNet. Compared to lightweight methods, WSERNet provides substantial accuracy improvements (6.64\% over MSNet, 8.82\% over SSwsrNet) with reasonable computational overhead increases. The joint optimization approach adds only 84,000 parameters and 1.6M FLOPs to WSRNet while providing 2.53\% accuracy improvement, demonstrating excellent cost-benefit trade-off.

\begin{figure}
\centering
\includegraphics[width=0.45\textwidth]{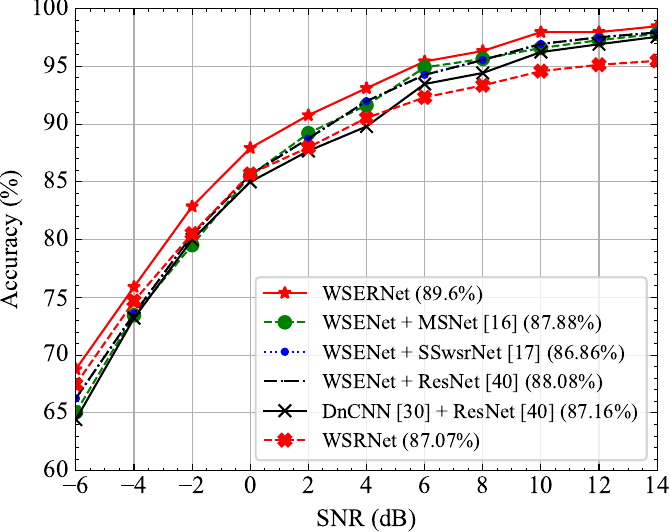}
\caption{Performance comparison with the state-of-the-art models including VGG \cite{o2016convolutional}, ResNet \cite{o2018over}, MSNet \cite{zhang2021novel}, and SSwsrNet \cite{zhang2024sswsrnet} under centralized conditions with WSENet.}
\label{fig:center_ser}
\end{figure}

To demonstrate the effectiveness of the proposed WSENet, we present the accuracy comparison with the state-of-the-art models including VGG \cite{o2016convolutional}, ResNet \cite{o2018over}, MSNet \cite{zhang2021novel}, and SSwsrNet \cite{zhang2024sswsrnet} under centralized conditions with WSENet. The results are shown in Fig. \ref{fig:center_ser}. As can be seen, the proposed WSERNet achieves the best performance, with an accuracy of $89.6$\%, which is higher than the state-of-the-art models with WSENet. Moreover, compared to WSRNet, the proposed WSERNet achieves an improvement of $2.53$\% in classification accuracy. 
Otherwise, MSNet \cite{zhang2021novel} ranks second with $87.88$\% accuracy, achieving a boost of $4.92$\% compared to the performance without WSENet.  
SSwsrNet \cite{zhang2024sswsrnet} and ResNet \cite{o2018over} show comparable performance, with accuracies of $86.86$\% and $88.08$\%, respectively, and improvements of $5.29$\% and $8.19$\% compared to the performance without WSENet. 
Additionally, we apply ResNet with DnCNN \cite{zhang2017beyond} for signal enhancement, achieving an accuracy of $87.16$\%, which is $0.92$\% lower than the performance with WSENet. 
The integration of WSENet with various models (MSNet, SSwsrNet, ResNet) demonstrates competitive performance, suggesting the effectiveness of WSENet in enhancing other architectures. 

\begin{figure*}
\centering
\begin{minipage}[b]{0.32\linewidth}
\centering
\subfigure[]{\includegraphics[width=\linewidth]{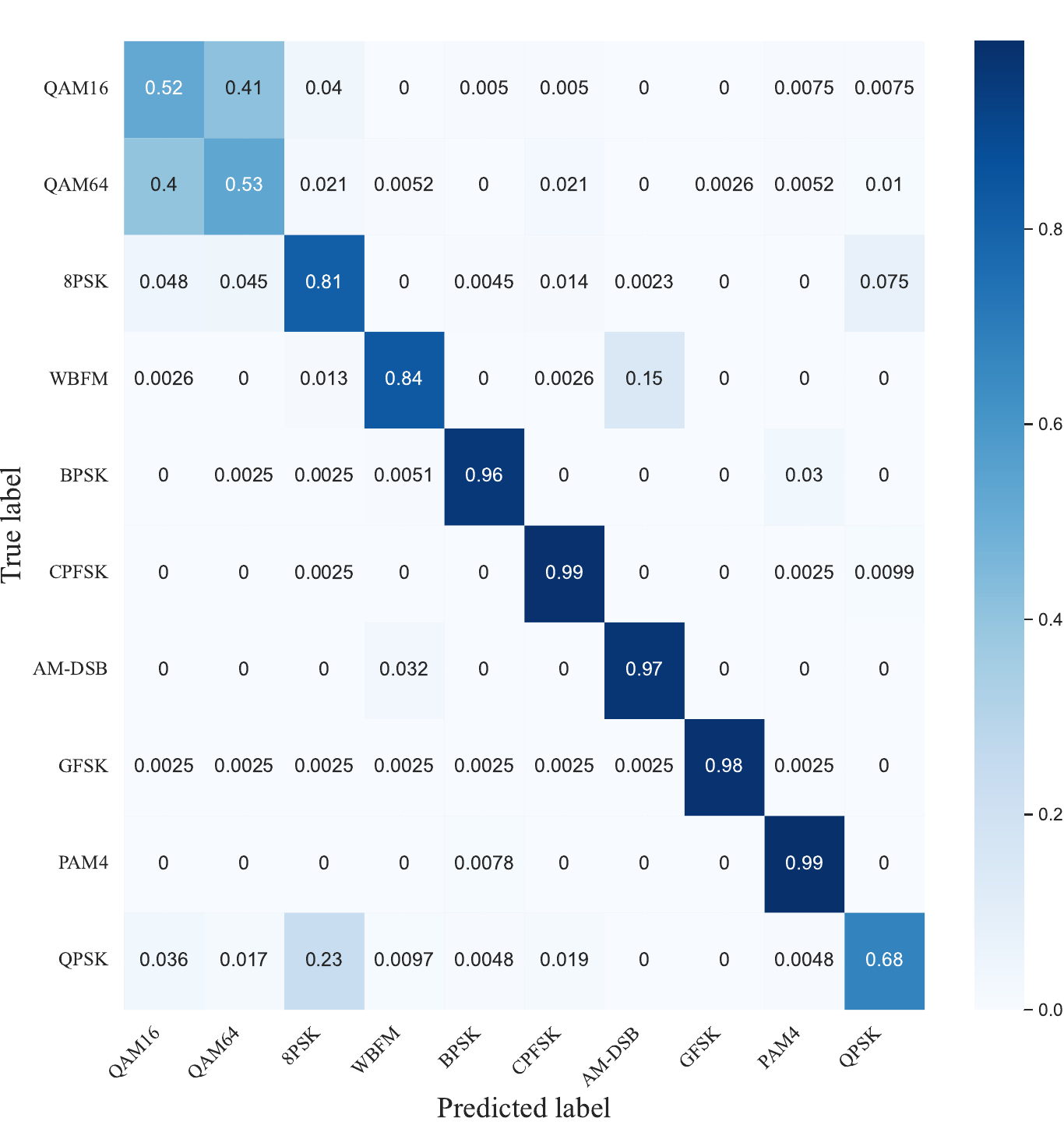}}
\end{minipage}
\hfill
\begin{minipage}[b]{0.32\linewidth}
\centering
\subfigure[]{\includegraphics[width=\linewidth]{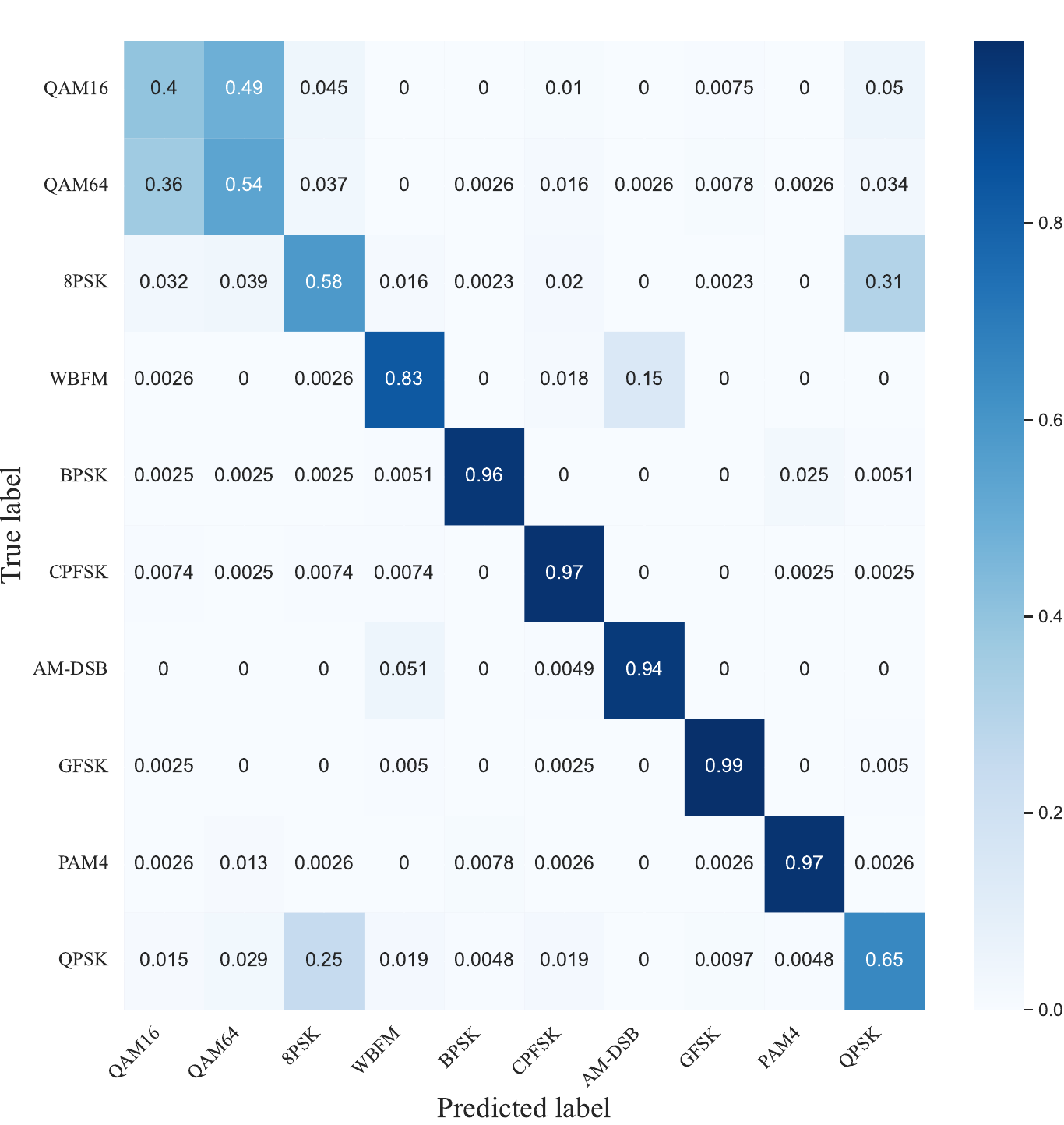}}
\end{minipage}
\hfill
\begin{minipage}[b]{0.32\linewidth}
    \centering
    \subfigure[]{\includegraphics[width=\linewidth]{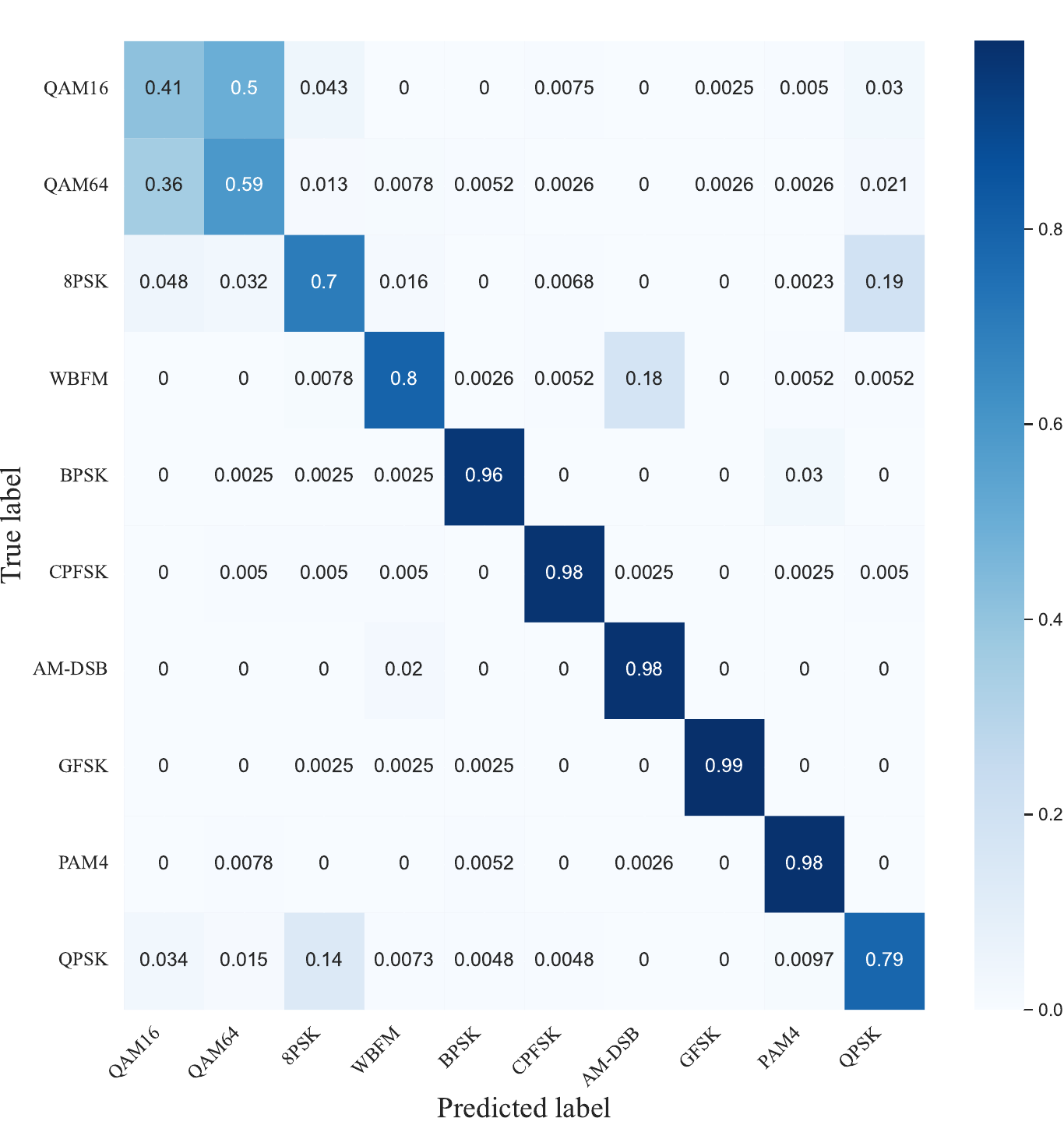}}
\end{minipage}

    
\begin{minipage}[b]{0.32\linewidth}
\centering
\subfigure[]{\includegraphics[width=\linewidth]{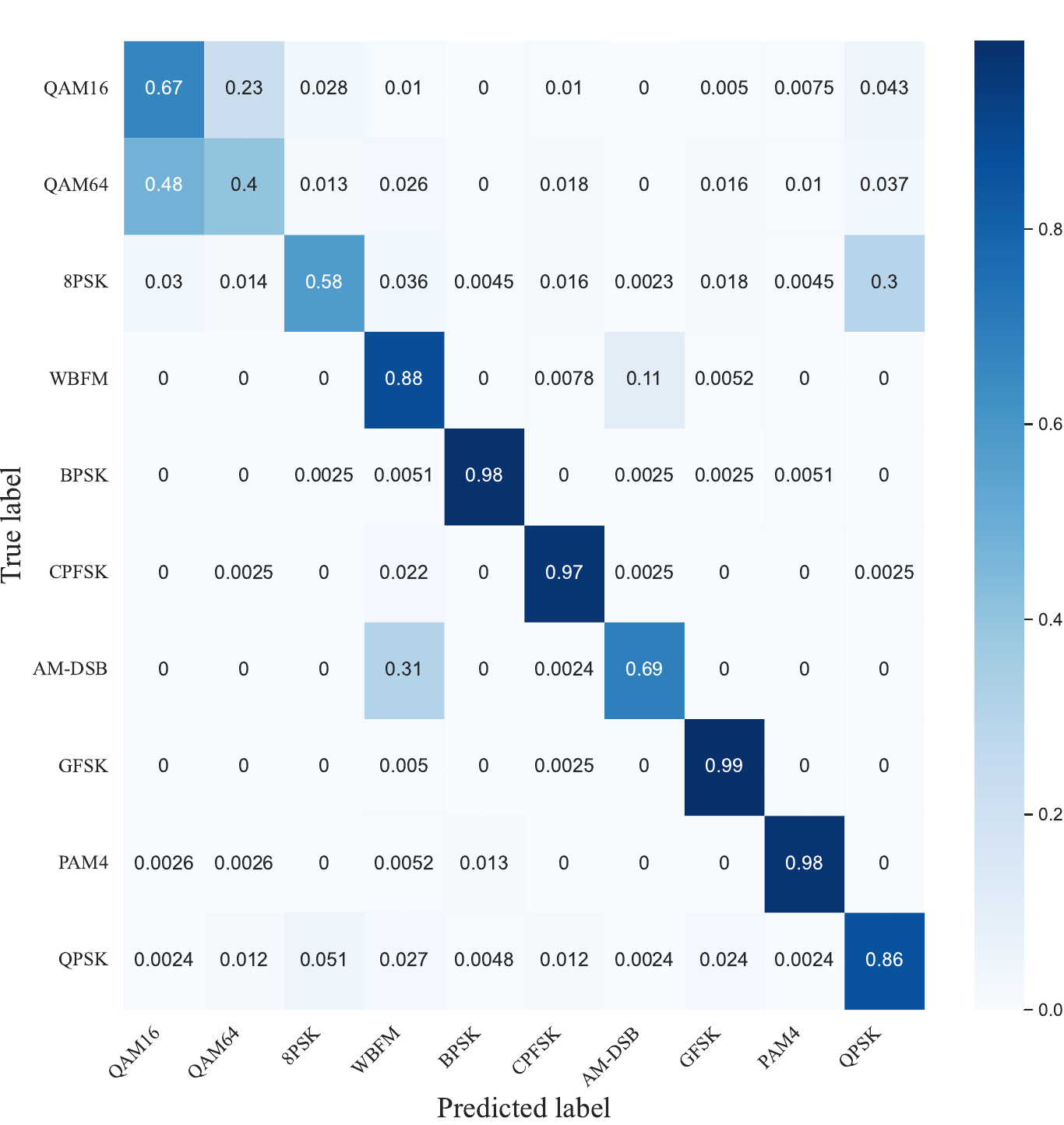}}
\end{minipage}
\hfill
\begin{minipage}[b]{0.32\linewidth}
\centering
\subfigure[]{\includegraphics[width=\linewidth]{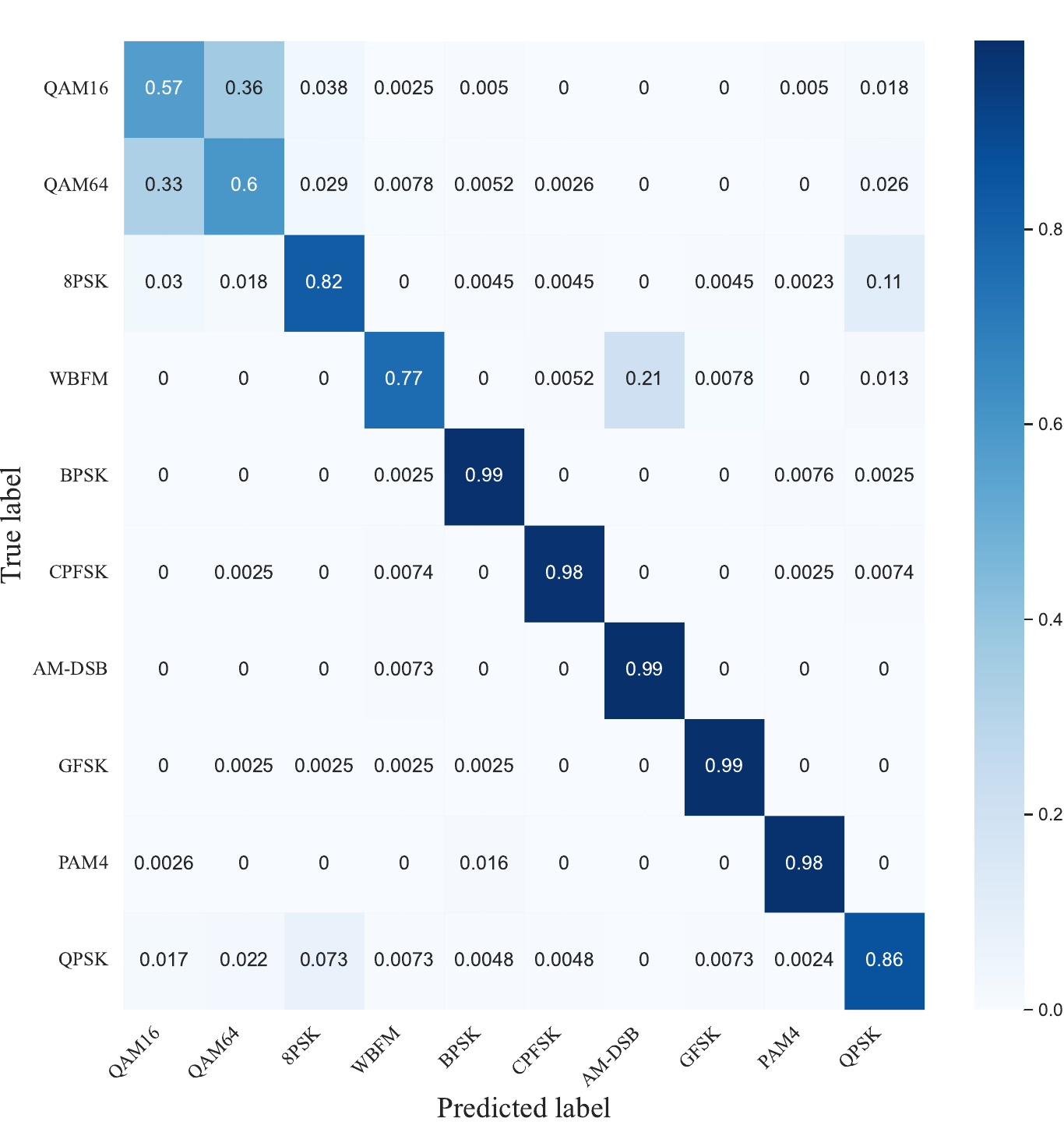}}
\end{minipage}
\hfill
\begin{minipage}[b]{0.32\linewidth}
\centering
\subfigure[]{\includegraphics[width=\linewidth]{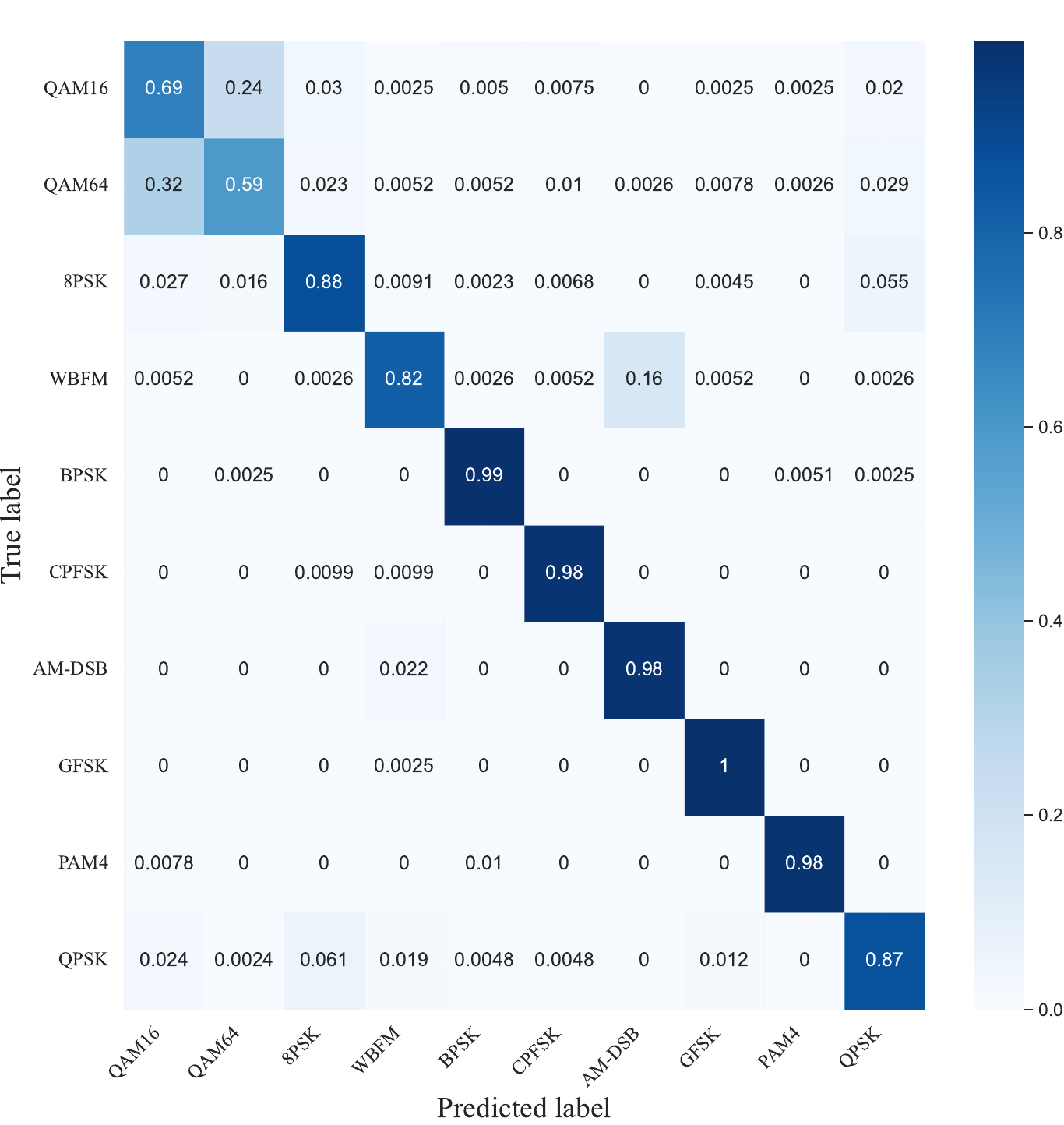}}
\end{minipage}
\caption{Confusion Matrix at $0$dB of (a) VGG \cite{o2016convolutional}, (b) ResNet \cite{o2018over}, (c) MSNet \cite{zhang2021novel}, (d) SSwsrNet \cite{zhang2024sswsrnet}, (e) WSRNet and (f) WSERNet under centralized conditions.}
\label{fig:confusion_matrix}
\end{figure*}

To better understand the performance of the proposed method, we visualize the confusion matrix of the proposed method with other state-of-the-art models including VGG \cite{o2016convolutional}, ResNet \cite{o2018over}, MSNet \cite{zhang2021novel}, and SSwsrNet \cite{zhang2024sswsrnet} under centralized conditions at $0$dB, which is shown in Fig. \ref{fig:confusion_matrix}. As can be seen, BPSK, CPFSK, GFSK, and PAM4 generally show high classification accuracy (larger than $0.95$) across most models. However, there is notable confusion between QAM16 and QAM64 in all models, with significant misclassification between these two classes. 
8PSK and QPSK also exhibit some degree of mutual misclassification across all models. WBFM and AM-DSB show some mutual confusion, particularly in models (a) through (d). 
The proposed WSERNet demonstrates robust performance in radio signal modulation classification at 0 dB SNR, achieving high accuracy ($>$0.98) for GFSK, BPSK, CPFSK, AM-DSB, and PAM4. It shows strong performance for 8PSK and QPSK (0.88 and 0.87 respectively), and reasonable accuracy for QAM64 and WBFM. The main challenge lies in distinguishing QAM16 and QAM64, which have the worst performance (0.69 and 0.59). Despite this, WSERNet achieves over 0.80 accuracy for 8 out of 10 modulation types, indicating strong overall performance across diverse modulation schemes. Compared to other models, it appears to handle the 8PSK/QPSK distinction better and shows improved QAM16 classification, although there's still room for enhancement in this area. These results suggest that WSERNet is a highly effective model for modulation classification in challenging noise conditions.

\subsection{Performance Evaluation under IID Distributed Conditions}
\begin{figure}
\centering
\includegraphics[width=0.45\textwidth]{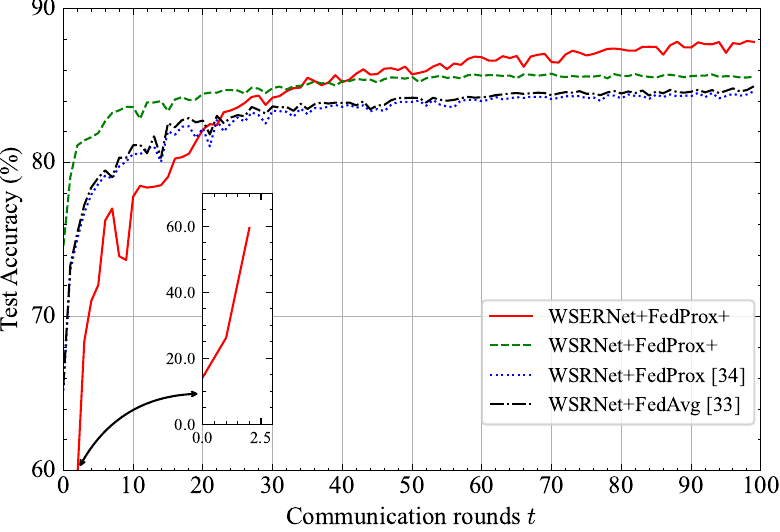}
\caption{Test accuracy with rounds under IID distributed training process of the proposed \texttt{FedProx+} algorithm compared with FedAvg \cite{mcmahan2017communication} and FedProx \cite{li2020federated}.}
\label{fig:iid_test_acc}
\end{figure}

To demonstrate the effectiveness of the proposed method under distributed conditions, we evaluate the performance of the proposed method \texttt{FedProx+}, FedAvg \cite{mcmahan2017communication} and FedProx \cite{li2020federated} under IID distributed conditions. 
Fig. \ref{fig:iid_test_acc} shows the test accuracy of different federated learning algorithms over multiple communication rounds for an IID-distributed training process. 
As can be seen, all methods show rapid improvement in the first 20-30 rounds, then gradual increases or stabilization. 
The proposed method WSERNet+FedProx+ achieves the highest final accuracy, reaching about $88\%$ after 100 rounds. 
Notably, WSERNet+FedProx+ starts with the lowest accuracy but overtakes other methods around round 30. This is mainly due to the joint optimization of signal enhancement and recognition tasks. 
WSRNet+FedProx+ performs second best, stabilizing at around 85\% accuracy.
WSRNet+FedProx and WSRNet+FedAvg show similar performance, both reaching about 84\% accuracy. 

\begin{figure}
\centering
\includegraphics[width=0.45\textwidth]{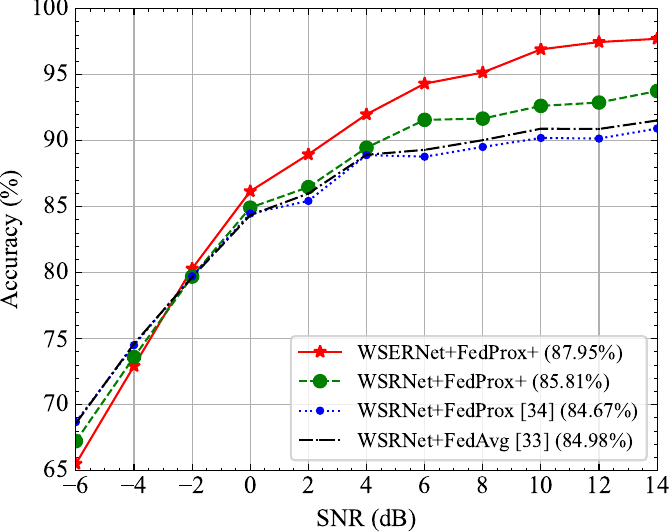}
\caption{Performance comparison under IID distributed conditions of the proposed \texttt{FedProx+} algorithm compared with FedAvg \cite{mcmahan2017communication} and FedProx \cite{li2020federated}.}
\label{fig:fed_iid_acc}
\end{figure}

To further present the performance of the proposed method under IID distributed conditions, we compare the accuracy of the proposed method WSERNet+FedProx+ with FedAvg \cite{mcmahan2017communication} and FedProx \cite{li2020federated} over multiple communication rounds across various SNR levels from $-6$ dB to $14$ dB, which is shown in Fig. \ref{fig:fed_iid_acc}. The proposed WSERNet+FedProx+ algorithm consistently outperforms the other methods, achieving the highest overall accuracy of 87.95\%. It shows particularly significant improvements at higher SNR levels, reaching nearly 98\% accuracy at 14 dB. WSRNet+FedProx+ follows as the second-best performer with 85.81\% accuracy, while WSRNet+FedAvg and WSRNet+FedProx show similar performance (84.98\% and 84.67\% respectively). All algorithms demonstrate improved accuracy as SNR increases, with performance differences becoming more pronounced at higher SNR levels. At low SNR (-6 dB), the algorithms perform similarly, but they start to diverge around -2 dB. This comparison clearly illustrates the superior performance of the WSERNet+FedProx+ algorithm across various noise conditions, especially in environments with higher SNRs.

\begin{figure}
\centering
\includegraphics[width=0.45\textwidth]{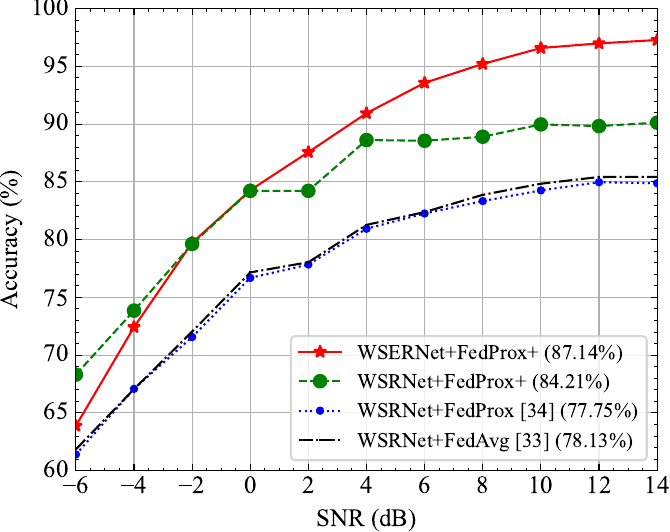}
\caption{Performance comparison under IID distributed conditions of the proposed \texttt{FedProx+} algorithm compared with FedAvg \cite{mcmahan2017communication} and FedProx \cite{li2020federated} using $25$ clients selected from $50$ distributed devices.}
\label{fig:fed_iid_acc_m25}
\end{figure}

Fig.\ref{fig:fed_iid_acc_m25} presents the performance of the proposed \texttt{FedProx+} under large-scale distributed conditions. The training data are split into $50$ clients in an IID manner, and $25$ devices are randomly selected for training. 
As shown in Fig. \ref{fig:fed_iid_acc_m25}, we compare four federated learning algorithms under IID conditions across various SNR levels (-6 dB to 14 dB). WSERNet+FedProx+ consistently outperforms the other methods, achieving the highest overall accuracy of 87.14\%. It shows significant improvements, especially at higher SNR levels, reaching nearly 98\% accuracy at 14 dB. WSRNet+FedProx+ follows with 84.21\% accuracy, while WSRNet+FedAvg and WSRNet+FedProx perform similarly (78.13\% and 77.75\% respectively). All algorithms improve as SNR increases, with performance gaps widening at higher SNRs. At low SNR (-6 dB), performances are similar, but they diverge as SNR improves. This comparison demonstrates the superior performance of WSERNet+FedProx+ across various noise conditions in a distributed network setting, particularly excelling in environments with higher SNRs.

\subsection{Performance Evaluation under Non-IID Distributed Conditions}

To demonstrate the effectiveness of the proposed method under non-IID distributed conditions, we evaluate the performance of the proposed method \texttt{FedProx+} under non-IID distributed conditions. 
Fig. \ref{fig:iid_test_acc} shows the test accuracy of different federated learning algorithms including FedAvg \cite{mcmahan2017communication} and FedProx \cite{li2020federated} over multiple communication rounds for a non-IID distributed training process. 
As can be seen, WSERNet+FedProx+ consistently outperforms the other methods, achieving the highest accuracy of around 80\% by the end of the training. WSRNet+FedProx+  shows the second-best performance, reaching about 75\% accuracy. WSRNet+FedAvg and WSRNet+FedProx  demonstrate similar performance patterns, both stabilizing around 70\% accuracy. All algorithms show rapid improvement in the initial rounds, with WSERNet+FedProx+ and WSRNet+FedProx+ displaying faster convergence. The graph illustrates that the proposed FedProx+ algorithm, especially when combined with WSERNet, offers superior performance in non-IID distributed settings, maintaining its advantage throughout the training process and achieving higher final accuracy compared to traditional FedAvg and FedProx algorithms.

\begin{figure}
\centering
\includegraphics[width=0.45\textwidth]{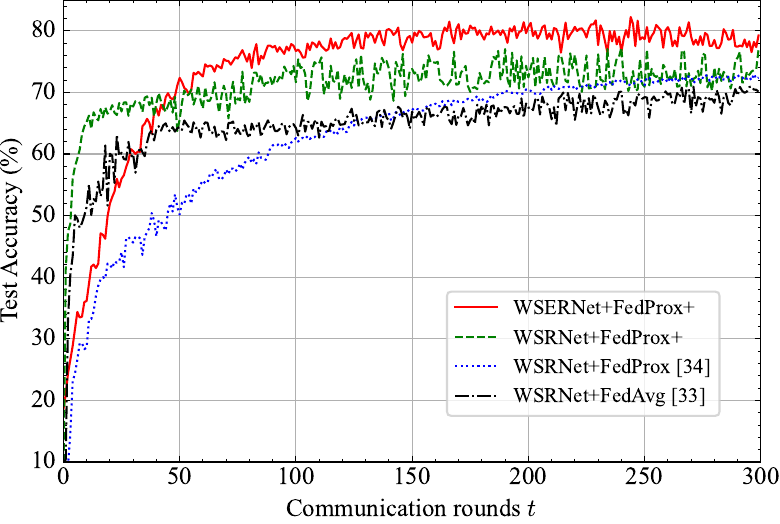}
\caption{Test accuracy with rounds under Non-IID distributed training process of the proposed \texttt{FedProx+} algorithm compared with FedAvg \cite{mcmahan2017communication} and FedProx \cite{li2020federated}, using $50$ clients selected from $100$ distributed devices.}
\label{fig:niid_test_acc}
\end{figure}

\begin{figure}
\centering
\includegraphics[width=0.45\textwidth]{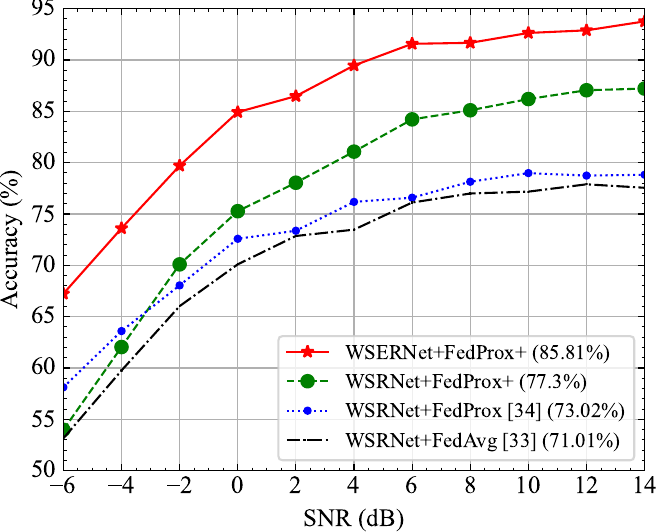}
\caption{Performance comparison under non-IID distributed conditions of the proposed \texttt{FedProx+} algorithm compared with FedAvg \cite{mcmahan2017communication} and FedProx \cite{li2020federated} using $50$ clients selected from $100$ distributed devices.}
\label{fig:fed_niid_acc}
\end{figure}

Fig.\ref{fig:fed_niid_acc} compares the performance of four federated learning algorithms under non-IID distributed conditions across various SNR levels, using 50 clients selected from 100 distributed devices. The WSERNet+FedProx+ algorithm consistently outperforms the other methods, achieving the highest overall accuracy of 85.81\%. It shows superior performance across all SNR levels, with the gap widening as SNR increases. WSRNet+FedProx+ is the second-best performer with 77.3\% accuracy, while WSRNet+FedProx and WSRNet+FedAvg show similar performance (73.02\% and 71.01\% respectively). All algorithms demonstrate improved accuracy as SNR increases from -6 dB to 14 dB, but WSERNet+FedProx+ shows the most significant improvement, reaching over 90\% accuracy at higher SNR levels. This comparison clearly illustrates the effectiveness of the proposed WSERNet+FedProx+ algorithm in non-IID distributed settings, particularly in environments with varying signal qualities, outperforming traditional FedAvg and FedProx algorithms across the entire SNR range.

\begin{figure}
\centering
\includegraphics[width=0.45\textwidth]{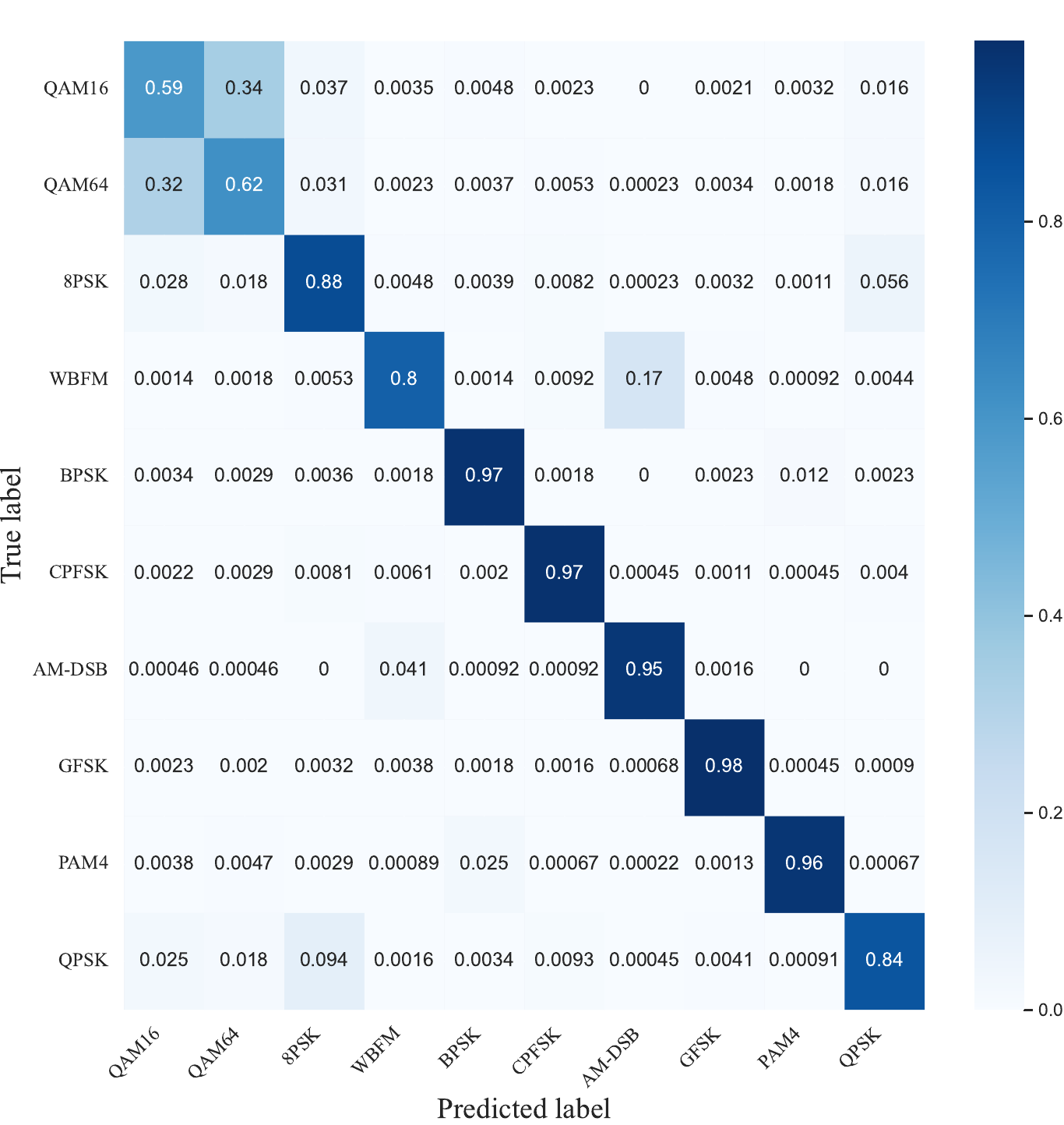}
\caption{Confusion matrix of the proposed WSERNet with \texttt{FedProx+} using $50$ clients selected from $100$ distributed devices.}
\label{fig:fed_niid_acc_confusion}
\end{figure}

\begin{figure}
\centering
\includegraphics[width=0.45\textwidth]{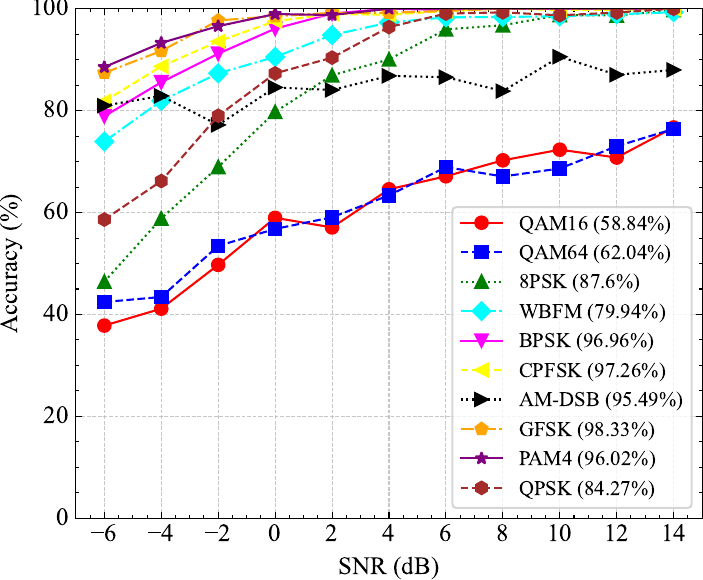}
\caption{Performance of different classes of the proposed WSERNet with \texttt{FedProx+}using $50$ clients selected from $100$ distributed devices.}
\label{fig:fed_niid_acc_class}
\end{figure}

To further present the performance of the proposed method under non-IID distributed conditions, we visualize the confusion matrix of the proposed method WSERNet+FedProx+, which is shown in Fig. \ref{fig:fed_niid_acc_confusion}. As can be seen, GFSK, BPSK, and CPFSK have the highest accuracies (0.98, 0.97, and 0.97 respectively), while QAM16 and QAM64 show lower but still reasonable accuracies (0.59 and 0.62). There's some confusion between certain pairs of classes, such as QAM16 and QAM64, as well as WBFM and AM-DSB. 
Furthermore, we visualize the performance of different classes with different SNRs of the proposed WSERNet+FedProx+ algorithm under non-IID distributed conditions, which is shown in Fig. \ref{fig:fed_niid_acc_class}. 
This result demonstrates that the proposed WSERNet+FedProx+ algorithm achieves high accuracy across all classes, with the highest accuracy for GFSK, BPSK, and CPFSK. The accuracy of QAM16 and QAM64 is lower but still reasonable, while WBFM and AM-DSB show some mutual confusion. This visualization provides a detailed understanding of the performance of the proposed WSERNet+FedProx+ algorithm under non-IID distributed conditions, highlighting its robustness and effectiveness in challenging signal recognition tasks.

\subsection{Discussion}
Our current framework has several limitations that should be acknowledged. First, while we address data distribution heterogeneity (IID vs. non-IID), we assume uniform channel conditions across federated clients, whereas practical deployments involve diverse propagation environments and hardware configurations. Second, our evaluation relies on the simulated RML22 dataset with a unified channel model, which may not fully capture the complexity of real-world signals including non-Gaussian noise and hardware impairments. Third, although WSERNet is designed to be lightweight, we have not provided detailed computational complexity analysis or energy consumption evaluation for resource-constrained edge devices. Finally, our federated learning framework does not explicitly address security considerations such as model poisoning attacks or privacy leakage.

Several promising research directions emerge from this work. First, developing channel-aware federated learning algorithms that explicitly model and adapt to diverse channel conditions across clients represents a critical extension. Second, real-world validation with actual over-the-air signal captures in diverse environments would strengthen the practical applicability of our findings. Third, integrating enhanced privacy protection mechanisms such as differential privacy and secure aggregation protocols would address security concerns in federated WSER systems. Fourth, extending the framework to handle multi-modal signal processing (\emph{e.g.}, radar, satellite communications) and other wireless signal recognition tasks such as wireless technology identification would broaden its applicability in various application scenarios \cite{zhuang2019sdn ,li2025digital}. Finally, edge computing optimization through model compression and quantization techniques would facilitate deployment on resource-constrained devices.

\section{Conclusion}
\label{sec:conclusion}
In this paper, we propose a novel distributed multi-task learning framework for joint wireless signal enhancement and recognition using federated learning. The proposed framework consists of a wireless signal enhancement and recognition network (WSERNet) and an enhanced FedProx algorithm \texttt{FedProx+}. The WSERNet is designed to be efficient while lightweight, making it suitable for deployment on resource-constrained edge devices. The proposed \texttt{FedProx+} algorithm incorporates client-specific and dynamically adjusted proximal term weights to handle heterogeneous data distributions across clients. The proposed framework is evaluated on a wide applied dataset, and the experimental results demonstrate that the proposed framework achieves superior performance compared to state-of-the-art methods for joint wireless signal enhancement and recognition. The proposed framework shows potential for extension to other multi-task learning-based signal processing applications.

\bibliographystyle{IEEEtran}
\bibliography{dsp_ref.bib}

\begin{thebibliography}{10}
\providecommand{\url}[1]{#1}
\csname url@samestyle\endcsname
\providecommand{\newblock}{\relax}
\providecommand{\bibinfo}[2]{#2}
\providecommand{\BIBentrySTDinterwordspacing}{\spaceskip=0pt\relax}
\providecommand{\BIBentryALTinterwordstretchfactor}{4}
\providecommand{\BIBentryALTinterwordspacing}{\spaceskip=\fontdimen2\font plus
\BIBentryALTinterwordstretchfactor\fontdimen3\font minus \fontdimen4\font\relax}
\providecommand{\BIBforeignlanguage}[2]{{%
\expandafter\ifx\csname l@#1\endcsname\relax
\typeout{** WARNING: IEEEtran.bst: No hyphenation pattern has been}%
\typeout{** loaded for the language `#1'. Using the pattern for}%
\typeout{** the default language instead.}%
\else
\language=\csname l@#1\endcsname
\fi
#2}}
\providecommand{\BIBdecl}{\relax}
\BIBdecl

\bibitem{wang2010advances}
B.~Wang and K.~R. Liu, ``Advances in cognitive radio networks: A survey,'' \emph{IEEE Journal of Selected Topics in Signal Processing}, vol.~5, no.~1, pp. 5--23, 2010.

\bibitem{zhang2025revolution}
H.~Zhang, F.~Zhou, H.~Du, Q.~Wu, and C.~Yuen, ``Revolution of wireless signal recognition for 6g: Recent advances, challenges and future directions,'' \emph{IEEE Communications Surveys \& Tutorials}, 2025.

\bibitem{dobre2007survey}
O.~A. Dobre, A.~Abdi, Y.~Bar-Ness, and W.~Su, ``Survey of automatic modulation classification techniques: classical approaches and new trends,'' \emph{IET communications}, vol.~1, no.~2, pp. 137--156, 2007.

\bibitem{huan1995likelihood}
C.-Y. Huan and A.~Polydoros, ``Likelihood methods for {MPSK} modulation classification,'' \emph{IEEE Transactions on Communications}, vol.~43, no. 2/3/4, pp. 1493--1504, 1995.

\bibitem{panagiotou2000likelihood}
P.~Panagiotou, A.~Anastasopoulos, and A.~Polydoros, ``Likelihood ratio tests for modulation classification,'' in \emph{MILCOM 2000 Proceedings. 21st Century Military Communications. Architectures and Technologies for Information Superiority (Cat. No. 00CH37155)}, vol.~2.\hskip 1em plus 0.5em minus 0.4em\relax IEEE, 2000, pp. 670--674.

\bibitem{hameed2009likelihood}
F.~Hameed, O.~A. Dobre, and D.~C. Popescu, ``On the likelihood-based approach to modulation classification,'' \emph{IEEE Transactions on Wireless Communications}, vol.~8, no.~12, pp. 5884--5892, 2009.

\bibitem{nandi1998algorithms}
A.~K. Nandi and E.~E. Azzouz, ``Algorithms for automatic modulation recognition of communication signals,'' \emph{IEEE Transactions on Communications}, vol.~46, no.~4, pp. 431--436, 1998.

\bibitem{li2021modulation}
T.~Li, Y.~Li, and O.~A. Dobre, ``Modulation classification based on fourth-order cumulants of superposed signal in {NOMA} systems,'' \emph{IEEE Transactions on Information Forensics and Security}, vol.~16, pp. 2885--2897, 2021.

\bibitem{yuan2004modulation}
J.~Yuan, Z.~Zhao-Yang, and Q.~Pei-Liang, ``Modulation classification of communication signals,'' in \emph{IEEE MILCOM 2004. Military Communications Conference, 2004.}, vol.~3.\hskip 1em plus 0.5em minus 0.4em\relax IEEE, 2004, pp. 1470--1476.

\bibitem{ramkumar2009automatic}
B.~Ramkumar, ``Automatic modulation classification for cognitive radios using cyclic feature detection,'' \emph{IEEE Circuits and Systems Magazine}, vol.~9, no.~2, pp. 27--45, 2009.

\bibitem{zhu2010augmented}
Z.~Zhu, M.~W. Aslam, and A.~K. Nandi, ``Augmented genetic programming for automatic digital modulation classification,'' in \emph{2010 IEEE International Workshop on Machine Learning for Signal Processing}.\hskip 1em plus 0.5em minus 0.4em\relax IEEE, 2010, pp. 391--396.

\bibitem{luan2022automatic}
S.~Luan, Y.~Gao, W.~Chen, N.~Yu, and Z.~Zhang, ``Automatic modulation classification: Decision tree based on error entropy and global-local feature-coupling network under mixed noise and fading channels,'' \emph{IEEE Wireless Communications Letters}, vol.~11, no.~8, pp. 1703--1707, 2022.

\bibitem{wang2009algorithm}
L.-X. Wang, Y.-J. Ren, and R.-H. Zhang, ``Algorithm of digital modulation recognition based on support vector machines,'' in \emph{2009 International Conference on Machine Learning and Cybernetics}, vol.~2.\hskip 1em plus 0.5em minus 0.4em\relax IEEE, 2009, pp. 980--983.

\bibitem{triantafyllakis2017phasma}
K.~Triantafyllakis, M.~Surligas, G.~Vardakis, and S.~Papadakis, ``Phasma: an automatic modulation classification system based on random forest,'' in \emph{2017 IEEE International Symposium on Dynamic Spectrum Access Networks (DySPAN)}.\hskip 1em plus 0.5em minus 0.4em\relax IEEE, 2017, pp. 1--3.

\bibitem{niazmand2025joint}
V.~Niazmand and Q.~Ye, ``Joint task offloading, dnn pruning, and computing resource allocation for fault detection with dynamic constraints in industrial iot,'' \emph{IEEE Transactions on Cognitive Communications and Networking}, 2025.

\bibitem{zhang2021novel}
H.~Zhang, F.~Zhou, Q.~Wu, W.~Wu, and R.~Q. Hu, ``A novel automatic modulation classification scheme based on multi-scale networks,'' \emph{IEEE Transactions on Cognitive Communications and Networking}, vol.~8, no.~1, pp. 97--110, 2021.

\bibitem{zhang2024sswsrnet}
H.~Zhang, F.~Zhou, Q.~Wu, and N.~Al-Dhahir, ``Sswsrnet: A semi-supervised few-shot learning framework for wireless signal recognition,'' \emph{IEEE Transactions on Communications}, 2024.

\bibitem{yuan2021multiscale}
L.~Yuan, H.~Zhang, M.~Xu, F.~Zhou, and Q.~Wu, ``A multiscale cnn framework for wireless technique classification in internet of things,'' \emph{IEEE Internet of Things Journal}, vol.~9, no.~12, pp. 10\,366--10\,367, 2021.

\bibitem{ding2022data}
R.~Ding, H.~Zhang, F.~Zhou, Q.~Wu, and Z.~Han, ``Data-and-knowledge dual-driven automatic modulation recognition for wireless communication networks,'' in \emph{ICC 2022-IEEE International Conference on Communications}.\hskip 1em plus 0.5em minus 0.4em\relax IEEE, 2022, pp. 1962--1967.

\bibitem{huang2020automatic}
S.~Huang, R.~Dai, J.~Huang, Y.~Yao, Y.~Gao, F.~Ning, and Z.~Feng, ``Automatic modulation classification using gated recurrent residual network,'' \emph{IEEE Internet of Things Journal}, vol.~7, no.~8, pp. 7795--7807, 2020.

\bibitem{wang2020transfer}
Y.~Wang, G.~Gui, H.~Gacanin, T.~Ohtsuki, H.~Sari, and F.~Adachi, ``Transfer learning for semi-supervised automatic modulation classification in {ZF-MIMO} systems,'' \emph{IEEE Journal on Emerging and Selected Topics in Circuits and Systems}, vol.~10, no.~2, pp. 231--239, 2020.

\bibitem{zhang2023frequency}
D.~Zhang, Y.~Lu, Y.~Li, W.~Ding, B.~Zhang, and J.~Xiao, ``Frequency learning attention networks based on deep learning for automatic modulation classification in wireless communication,'' \emph{Pattern Recognition}, vol. 137, p. 109345, 2023.

\bibitem{zhang2024few}
X.~Zhang, Y.~Wang, H.~Huang, Y.~Lin, H.~Zhao, and G.~Gui, ``Few-shot automatic modulation classification using architecture search and knowledge transfer in radar-communication coexistence scenarios,'' \emph{IEEE Internet of Things Journal}, 2024.

\bibitem{xu2024distributed}
Y.~Xu, E.~G. Larsson, E.~A. Jorswieck, X.~Li, S.~Jin, and T.-H. Chang, ``Distributed signal processing for extremely large-scale antenna array systems: State-of-the-art and future directions,'' \emph{arXiv preprint arXiv:2407.16121}, 2024.

\bibitem{qian2022distributed}
L.~Qian, P.~Yang, M.~Xiao, O.~A. Dobre, M.~Di~Renzo, J.~Li, Z.~Han, Q.~Yi, and J.~Zhao, ``Distributed learning for wireless communications: Methods, applications and challenges,'' \emph{IEEE Journal of Selected Topics in Signal Processing}, vol.~16, no.~3, pp. 326--342, 2022.

\bibitem{gao2009denoising}
J.~Gao, H.~Sultan, J.~Hu, and W.-W. Tung, ``Denoising nonlinear time series by adaptive filtering and wavelet shrinkage: a comparison,'' \emph{IEEE Signal Processing Letters}, vol.~17, no.~3, pp. 237--240, 2009.

\bibitem{tracey2012nonlocal}
B.~H. Tracey and E.~L. Miller, ``Nonlocal means denoising of ecg signals,'' \emph{IEEE Transactions on Biomedical Engineering}, vol.~59, no.~9, pp. 2383--2386, 2012.

\bibitem{sendur2002bivariate}
L.~Sendur and I.~W. Selesnick, ``Bivariate shrinkage functions for wavelet-based denoising exploiting interscale dependency,'' \emph{IEEE Transactions on Signal Processing}, vol.~50, no.~11, pp. 2744--2756, 2002.

\bibitem{chiron2014efficient}
L.~Chiron, M.~A. van Agthoven, B.~Kieffer, C.~Rolando, and M.-A. Delsuc, ``Efficient denoising algorithms for large experimental datasets and their applications in fourier transform ion cyclotron resonance mass spectrometry,'' \emph{Proceedings of the National Academy of Sciences}, vol. 111, no.~4, pp. 1385--1390, 2014.

\bibitem{zhang2017beyond}
K.~Zhang, W.~Zuo, Y.~Chen, D.~Meng, and L.~Zhang, ``Beyond a gaussian denoiser: Residual learning of deep cnn for image denoising,'' \emph{IEEE Transactions on Image Processing}, vol.~26, no.~7, pp. 3142--3155, 2017.

\bibitem{liu2017deep}
X.~Liu, D.~Yang, and A.~El~Gamal, ``Deep neural network architectures for modulation classification,'' in \emph{2017 51st Asilomar Conference on Signals, Systems, and Computers}.\hskip 1em plus 0.5em minus 0.4em\relax IEEE, 2017, pp. 915--919.

\bibitem{zhang2020automatic}
Z.~Zhang, H.~Luo, C.~Wang, C.~Gan, and Y.~Xiang, ``Automatic modulation classification using {CNN-LSTM} based dual-stream structure,'' \emph{IEEE Transactions on Vehicular Technology}, vol.~69, no.~11, pp. 13\,521--13\,531, 2020.

\bibitem{zhang2025federated}
H.~Zhang, F.~Zhou, W.~Wang, Q.~Wu, and C.~Yuen, ``A federated learning-based lightweight network with zero trust for uav authentication,'' \emph{IEEE Transactions on Information Forensics and Security}, vol.~20, pp. 7424--7437, 2025.

\bibitem{li2020federated}
T.~Li, A.~K. Sahu, M.~Zaheer, M.~Sanjabi, A.~Talwalkar, and V.~Smith, ``Federated optimization in heterogeneous networks,'' \emph{Proceedings of Machine Learning and Systems}, vol.~2, pp. 429--450, 2020.

\bibitem{sathyanarayanan2023rml22}
V.~Sathyanarayanan, P.~Gerstoft, and A.~El~Gamal, ``Rml22: Realistic dataset generation for wireless modulation classification,'' \emph{IEEE Transactions on Wireless Communications}, vol.~22, no.~11, pp. 7663--7675, 2023.

\bibitem{sandler2018mobilenetv2}
M.~Sandler, A.~Howard, M.~Zhu, A.~Zhmoginov, and L.-C. Chen, ``Mobilenetv2: Inverted residuals and linear bottlenecks,'' in \emph{Proceedings of the IEEE conference on computer vision and pattern recognition}, 2018, pp. 4510--4520.

\bibitem{tan2021efficientnetv2}
M.~Tan and Q.~Le, ``Efficientnetv2: Smaller models and faster training,'' in \emph{International Conference on Machine Learning}.\hskip 1em plus 0.5em minus 0.4em\relax PMLR, 2021, pp. 10\,096--10\,106.

\bibitem{o2016convolutional}
T.~J. O’Shea, J.~Corgan, and T.~C. Clancy, ``Convolutional radio modulation recognition networks,'' in \emph{International Conference on Engineering Applications of Neural Networks}.\hskip 1em plus 0.5em minus 0.4em\relax Springer, 2016, pp. 213--226.

\bibitem{paszke2019pytorch}
A.~Paszke, S.~Gross, F.~Massa, A.~Lerer, J.~Bradbury, G.~Chanan, T.~Killeen, Z.~Lin, N.~Gimelshein, L.~Antiga \emph{et~al.}, ``Pytorch: An imperative style, high-performance deep learning library,'' \emph{Advances in Neural Information Processing Systems}, vol.~32, 2019.

\bibitem{o2018over}
T.~J. O’Shea, T.~Roy, and T.~C. Clancy, ``Over-the-air deep learning based radio signal classification,'' \emph{IEEE Journal of Selected Topics in Signal Processing}, vol.~12, no.~1, pp. 168--179, 2018.

\bibitem{he2016deep}
K.~He, X.~Zhang, S.~Ren, and J.~Sun, ``Deep residual learning for image recognition,'' in \emph{Proceedings of the IEEE Conference on Computer Vision and Pattern Recognition}, 2016, pp. 770--778.

\bibitem{mcmahan2017communication}
B.~McMahan, E.~Moore, D.~Ramage, S.~Hampson, and B.~A. y~Arcas, ``Communication-efficient learning of deep networks from decentralized data,'' in \emph{Artificial Intelligence and Statistics}.\hskip 1em plus 0.5em minus 0.4em\relax PMLR, 2017, pp. 1273--1282.

\bibitem{zhuang2019sdn}
W.~Zhuang, Q.~Ye, F.~Lyu, N.~Cheng, and J.~Ren, ``Sdn/nfv-empowered future iov with enhanced communication, computing, and caching,'' \emph{Proceedings of the IEEE}, vol. 108, no.~2, pp. 274--291, 2019.

\bibitem{li2025digital}
Q.~Li, Q.~J. Ye, N.~Zhang, W.~Zhang, and F.~Hu, ``Digital-twin-enabled industrial iot: Vision, framework, and future directions,'' \emph{IEEE Wireless Communications}, 2025.

\end{thebibliography}


\begin{IEEEbiography}[{\includegraphics[width=1in,height=1.25in,clip,keepaspectratio]{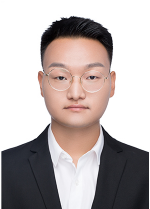}}]{Hao Zhang} (Member, IEEE) received the B.E. and M.Eng. from Nanchang University, China, in 2017 and 2020, respectively, and the Ph.D. degree from the College of Electronic and Information Engineering, Nanjing University of Aeronautics and Astronautics, Nanjing, China, in 2025. He is now a postdoctoral fellow with the College of Artificial Intelligence, Nanjing University of Aeronautics and Astronautics, Nanjing, China. He was a visiting Ph.D. student at the School of Electrical \& Electronic Engineering, Nanyang Technological University, Singapore, in 2024. His research interests focus on deep learning, foundation models, wireless communication, wireless signal processing, and spectrum cognition.
\end{IEEEbiography}

\vspace{-4em}

\begin{IEEEbiography}[{\includegraphics[width=1in,height=1.25in,clip,keepaspectratio]{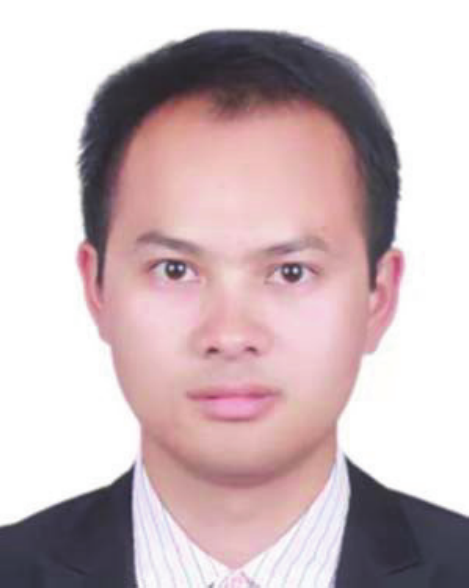}}]{Fuhui Zhou} (Senior member, IEEE) s currently a Full Professor with Nanjing University of Aeronautics and Astronautics, Nanjing, China, where he is also with the Key Laboratory of Dynamic Cognitive System  of  Electromagnetic  Spectrum  Space.  His research interests include cognitive radio, cognitive intelligence, knowledge graph, edge computing, and resource allocation.

Prof.  Zhou  has  published   over  200  papers   in internationally renowned journals and conferences in the field of communications. He has been selected for  1  ESI  hot  paper  and   13  ESI  highly  cited  papers.  He  has  received  4 Best Paper Awards at international conferences such as IEEE Globecom and IEEE  ICC.  He  was  awarded  as  2021  Most  Cited  Chinese  Researchers  by Elsevier,  Stanford  World’s  Top  2\%  Scientists,  IEEE  ComSoc  Asia-Pacific Outstanding Young Researcher and Young Elite Scientist Award of China and URSI GASS Young Scientist. He serves as an Editor of IEEE Transactions on  communication,  IEEE  Systems  Journal,  IEEE  Wireless  Communication Letters, IEEE Access and Physical Communications.
\end{IEEEbiography}
\vspace{-4em}

\begin{IEEEbiography}[{\includegraphics[width=1in,height=1.25in,clip,keepaspectratio]{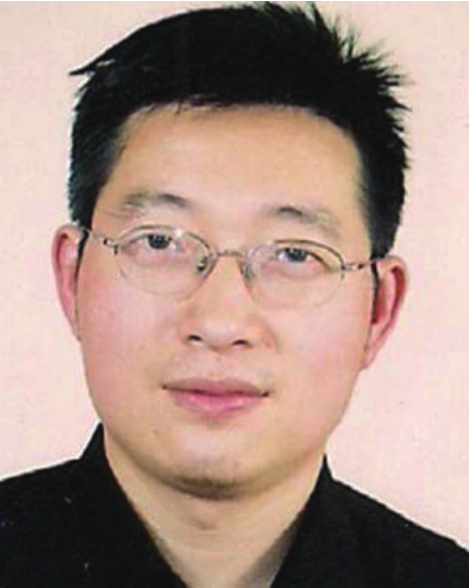}}]{Qihui Wu}
(Fellow, IEEE) received the B.S. degree in communications engineering, the M.S. and Ph.D. degrees in communications and information systems from the Institute of Communications Engineering, Nanjing, China, in 1994, 1997, and 2000, respectively. From 2003 to 2005, he was a Postdoctoral Research Associate with Southeast University, Nanjing, China. From 2005 to 2007, he was an Associate Professor with the College of Communications Engineering, PLA University of Science and Technology, Nanjing, China, where he was a Full Professor from 2008 to 2016. Since May 2016, he has been a Full Professor with the College of Electronic and Information Engineering, Nanjing University of Aeronautics and Astronautics, Nanjing, China. From March 2011 to September 2011, he was an Advanced Visiting Scholar with the Stevens Institute of Technology, Hoboken, USA. His current research interests span the areas of wireless communications and statistical signal processing, with emphasis on system design of software defined radio, cognitive radio, and smart radio.
\end{IEEEbiography}

\vspace{-4em}

\begin{IEEEbiography}[{\includegraphics[width=1in,height=1.25in,clip,keepaspectratio]{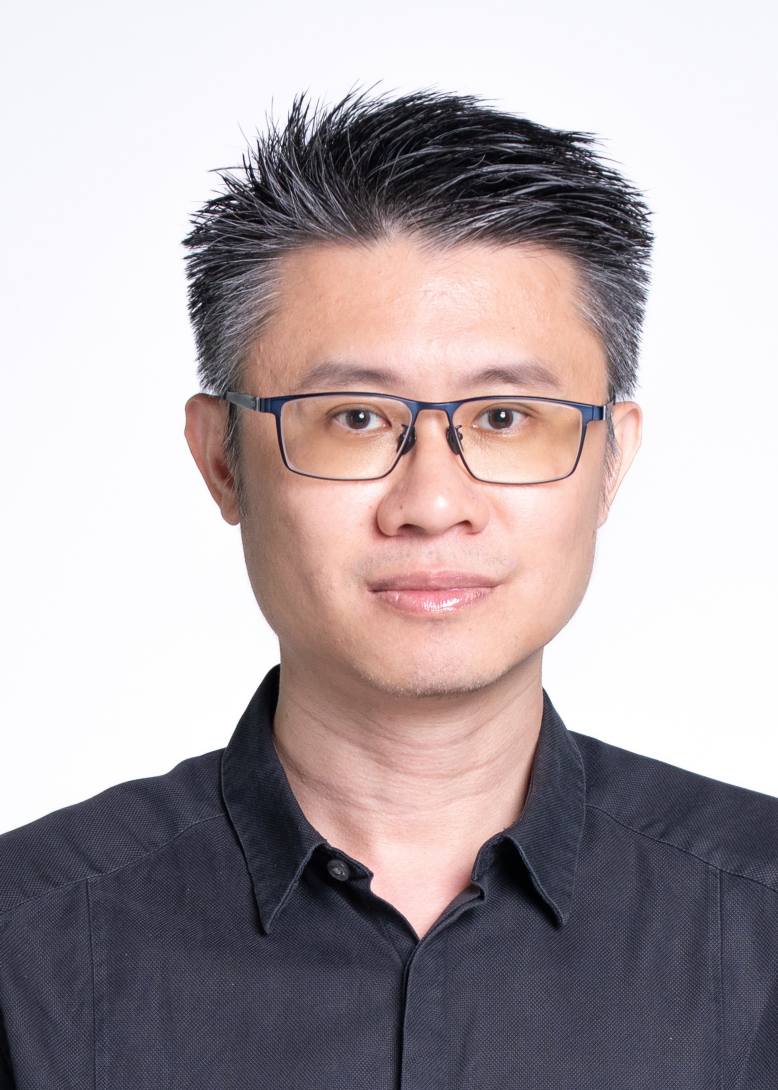}}]{Chau Yuen}
(S'02-M'06-SM'12-F'21) received the B.Eng. and Ph.D. degrees from Nanyang Technological University, Singapore, in 2000 and 2004, respectively. He was a Post-Doctoral Fellow with Lucent Technologies Bell Labs, Murray Hill, in 2005. From 2006 to 2010, he was with the Institute for Infocomm Research, Singapore. From 2010 to 2023, he was with the Engineering Product Development Pillar, Singapore University of Technology and Design. Since 2023, he has been with the School of Electrical and Electronic Engineering, Nanyang Technological University, currently he is Provost’s Chair in Wireless Communications, Assistant Dean in Graduate College, and Cluster Director for Sustainable Built Environment at ER@IN.
 
Dr. Yuen received IEEE Communications Society Leonard G. Abraham Prize (2024), IEEE Communications Society Best Tutorial Paper Award (2024), IEEE Communications Society Fred W. Ellersick Prize (2023), IEEE Marconi Prize Paper Award in Wireless Communications (2021), IEEE APB Outstanding Paper Award (2023), and EURASIP Best Paper Award for JOURNAL ON WIRELESS COMMUNICATIONS AND NETWORKING (2021).
 
Dr Yuen current serves as an Editor-in-Chief for Springer Nature Computer Science, Editor for IEEE TRANSACTIONS ON VEHICULAR TECHNOLOGY, IEEE TRANSACTIONS ON NEURAL NETWORKS AND LEARNING SYSTEMS, and IEEE TRANSACTIONS ON NETWORK SCIENCE AND ENGINEERING, where he was awarded as IEEE TNSE Excellent Editor Award 2024 and 2022, and Top Associate Editor for TVT from 2009 to 2015. He also served as the guest editor for several special issues, including IEEE JOURNAL ON SELECTED AREAS IN COMMUNICATIONS, IEEE WIRELESS COMMUNICATIONS MAGAZINE, IEEE COMMUNICATIONS MAGAZINE, IEEE VEHICULAR TECHNOLOGY MAGAZINE, IEEE TRANSACTIONS ON COGNITIVE COMMUNICATIONS AND NETWORKING, and ELSEVIER APPLIED ENERGY.
 
He is listed as Top 2\% Scientists by Stanford University, and also a Highly Cited Researcher by Clarivate Web of Science from 2022. He has 4 US patents and published over 500 research papers at international journals.
\end{IEEEbiography}

\end{document}